\documentclass[sigconf,authorversion,nonacm]{acmart}
\usepackage{preprint_header}
\AtBeginDocument{%
  }

\copyrightyear{2023} 
\acmYear{2023} 
\setcopyright{acmlicensed}\acmConference[SIGIR '23]{Proceedings of the 46th International ACM SIGIR Conference on Research and Development in Information Retrieval}{July 23--27, 2023}{Taipei, Taiwan}
\acmBooktitle{Proceedings of the 46th International ACM SIGIR Conference on Research and Development in Information Retrieval (SIGIR '23), July 23--27, 2023, Taipei, Taiwan}
\acmPrice{15.00}
\acmDOI{10.1145/3539618.3591882}
\acmISBN{978-1-4503-9408-6/23/07}

\usepackage{makecell}
\usepackage{tabularx}
\usepackage{xspace}
\usepackage{cleveref}
\usepackage{threeparttable}
\usepackage{graphicx}
\usepackage{hyperref}
\usepackage{multirow}
\usepackage{subcaption}
\usepackage{float}
\usepackage{flafter} 
\usepackage{balance}

\newcommand{\mbibfull}{Media Bias Identification Benchmark\xspace}
\newcommand{\mbib}{MBIB\xspace}

\begin{document}

\title{Introducing MBIB - the first Media Bias Identification Benchmark Task and Dataset Collection}

\author{Martin Wessel}
\orcid{0000-0002-8152-4038}
\affiliation{%
  \institution{University of Konstanz}
  \city{Konstanz}
  \country{Germany}
  \postcode{43017-6221}
} \email{m.wessel@media-bias-research.org}

\author{Tomáš Horych}
\affiliation{%
  \institution{Czech Technical University}
  \city{Prague}
  \country{Czech Republic}}
\email{t.horych@media-bias-research.org}

\author{Terry Ruas}
\affiliation{%
 \institution{University of Göttingen}
 \city{Göttingen}
 \country{Germany}}
\email{ruas@uni-goettingen.de}

\author{Akiko Aizawa}
\affiliation{%
 \institution{National Institute of Informatics}
 \city{Tokyo}
 \country{Japan}}
\email{aizawa@nii.ac.jp}

\author{Bela Gipp}
\affiliation{%
 \institution{University of Göttingen}
 \city{Göttingen}
 \country{Germany}}
\email{gipp@uni-goettingen.de}

\author{Timo Spinde}
\affiliation{%
  \institution{University of Göttingen}
  \city{Göttingen}
  \country{Germany}}
\email{t.spinde@media-bias-research.org}

\renewcommand{\shortauthors}{Wessel et al.}

\begin{abstract}

Although media bias detection is a complex multi-task problem, there is, to date, no unified benchmark grouping these evaluation tasks.
We introduce the \mbibfull (\mbib), a comprehensive benchmark that groups different types of media bias (e.g., linguistic, cognitive, political) under a common framework to test how prospective detection techniques generalize.
After reviewing 115 datasets, we select nine tasks and carefully propose 22 associated datasets for evaluating media bias detection techniques. 
We evaluate \mbib using state-of-the-art Transformer techniques (e.g., T5, BART).
Our results suggest that while hate speech, racial bias, and gender bias are easier to detect, models struggle to handle certain bias types, e.g., cognitive and political bias.
However, our results show that no single technique can outperform all the others significantly.
We also find an uneven distribution of research interest and resource allocation to the individual tasks in media bias.
A unified benchmark encourages the development of more robust systems and shifts the current paradigm in media bias detection evaluation towards solutions that tackle not one but multiple media bias types simultaneously.

\end{abstract}




\maketitle
\thispagestyle{preprintbox}
\section{Introduction} \label{sec:introduction}
Media bias is often related to content favoring a particular viewpoint or ideology (e.g., political) \cite{recasensLinguisticModelsAnalyzing2013}.  
Such bias has been the focus of various research projects \cite{spinde-you-2021} and is generally defined and summarized by the term \textbf{media bias} \cite{spinde-how-2021}. 
Media bias can have various negative impacts, e.g., the distribution of false facts, affecting decision-making processes, and endangering the readers' trust in news \cite{gerberDoesMediaMatter2009}.
Accordingly, drawing attention to instances of media bias can have far-reaching benefits and counterbalance its impact \cite{baumer-testing-2015}. 
Although it may not be possible to eliminate media bias completely, acknowledging its presence helps educate readers about it and encourages journalists and publishers to evaluate their work impartially. 
Given the vast amount of digital information available, getting an overview of bias in various media outlets is only possible with automated solutions \cite{spinde-interdisciplinary-2021}. 
Therefore, automatic media bias detection experiences steadily growing research attention.

While substantial research exists for solving specific media bias detection tasks, e.g., gender bias in the news \cite{vanderpas2020}, they usually concentrate on identifying a single aspect within the media bias spectrum. 
To the best of our knowledge, no single uniform benchmark allows for the comparison of prominent models, and no overview of potential media bias tasks and subtasks exists.
The lack of a benchmark leads to some problems: (a) solutions are focused on specific subtasks and bias types, (b) no standardized comparison between models is possible, and (c) models in the domain are less likely to make use of multi-task learning, while media bias itself is a multi-task problem \cite{spinde-exploiting-2021}. 
A unified benchmark covering different bias aspects will, in the future, also allow the development of more robust systems.
Therefore, we propose the \mbibfull (\mbib) to introduce a challenging media bias task collection with associated datasets.
Our benchmark is composed of nine tasks and 22 associated datasets. 
As \mbib covers a wide range of media bias types (e.g., framing, political)\footnote{Full list in \Cref{tab:categories and datasets}.}, its solution provides a reliable proxy for evaluation covering the domain as entirely and socially relevant as possible.
To curate the 22 associated datasets, we conduct an extensive literature search, concluding with a list of 115 related datasets. 
In the following, we briefly summarize our contributions.

\begin{enumerate}
    \item The first media bias benchmark is composed of nine tasks and 22 datasets.
    \item A framework for evaluating models in a standardized way. 
    \item \mbib is publicly available at  \begin{center} \textbf{\url{https://github.com/Media-Bias-Group/MBIB}\footnote{To facilitate easy access, we also share our \mbib base corpus on huggingface so that it can be fetched directly. We provide detailed information on the usage of all our code and data within the GitHub repository.}\label{mbiburl}} \end{center}
\end{enumerate}

\section{Related Work} 
\subsection{Media bias} 
While media bias is generally often referred to as communicating about something ``in a prejudiced manner or with slanted viewpoint'' \cite{Lee_2021}, many definitions for media bias and its subtypes exist in the literature.
For instance, definitions of hate speech in the literature differ on whether and how strongly offensive language can constitute hate speech \cite{mathewHateXplainBenchmarkDataset2021a}. 
To name another example, within linguistic bias, some works rely stronger on traditional linguistic features, focusing on bias rather as an objective entity \cite{recasensLinguisticModelsAnalyzing2013}. 
Others use linguistic bias, or bias by word choice, as a majorly subjective concept \cite{spindeNeuralMediaBias2021}, or as a general word choice communicating stereotypes \cite{beukeboom_linguistic_2017}. 
In their literature review, \citet{spindeIntroducingMediaBias2022} show how the various definitions diverge in detail. 
They also introduce the media bias framework \cite{spindeIntroducingMediaBias2022}, a coherent overview of the current state of research on media bias from different perspectives, such as linguistic bias, text-level context bias, cognitive bias, group bias, and others. 
The authors show that media bias detection is a highly active research field and that transformer-based approaches have significantly improved the media bias classification task over recent years.  
However, most advances in media bias detection are still tied to single tasks (e.g., linguistic bias) and do not report on the generalizability of their results \cite{spindeNeuralMediaBias2021,fanPlainSightMedia2019, huguetcabotUsVsThem2021, liuRoBERTaRobustlyOptimized2019,liuPOLITICSPretrainingSamestory2022,wangLiarLiarPants2017a}. 
We give a full overview of existing datasets in \Cref{sec:buildingMbib}.
Overlapping and unclear definitions of media bias and its subtypes can lead authors to explore different biases under the same name. 
Current system evaluations are limited without a good overview of available definitions and datasets, and results are difficult to compare 
\cite{spindeIntroducingMediaBias2022}.\footnote{The low comparability is even more surprising given that many existing works highlight how important task awareness is in resource-scarce domains \cite{aribandi_ext5_2021} and in particular for media bias research \cite{krieger_domain-adaptive_2022, spindeNeuralMediaBias2021, spinde-towards-2021}.} 
The lack of a standardized evaluation procedure makes comparing and reproducing results even more challenging. 
Moreover, focusing on individual tasks does not allow for a holistic assessment of systems for media bias.
Developing systems that detect media bias as a whole will increase efficiency (as it reduces the need for multiple methods), allow for consistency (as it reduces the variation between evaluations), and lead to more comprehensive bias assessments.
By introducing \mbib, we aim to make such a development possible.

\subsection{Language Processing Benchmarks} \label{sec:benchmarks}
In other areas of natural language processing (outside of media bias), task benchmarks have shown how important proxy tasks can be when tackling complex problems.
For example, the General Language Understanding Evaluation (GLUE) benchmark \cite{wangGLUEMultiTaskBenchmark2019} decomposes natural language understanding into smaller tasks, formalizing the need for models with strong generalizability in the domain.
GLUE introduces various language understanding tasks, such as similarity, inference, and question-answering tasks. 
It also provides a standardized score that can be used to compare the performance of different NLP systems across different tasks.
An improved version of GLUE is SuperGLUE \cite{wangSuperGLUEStickierBenchmark2019}, which adds a broader variety of tasks\footnote{While the tasks in GLUE mainly involve sentence-level and word-level understanding, SuperGLUE tasks involve understanding and reasoning about more complex phenomena, such as temporal relations, coreference, and commonsense knowledge.} and also includes fewer training examples, making it more challenging to learn and generalize to new examples. 
The question answering included in SuperGLUE is increased in difficulty by requiring more contextual understanding and the combination of information from different parts of the text.
For both, SuperGLUE and GLUE, a task consists of a single labeled dataset.
Another benchmark follows a similar strategy for natural language processing systems, BIG-Bench \cite{srivastavaImitationGameQuantifying2022}. 
Opposed to those in existing benchmarks, media bias datasets are too small or only cover a particular aspect of a task. 
For example, they describe bias by word choice without a perspective on linguistic bias in total \cite{spindeNeuralMediaBias2021}. 

As existing benchmarks have successfully formalized a challenging problem by decomposing it into several defined subtasks, they can serve as inspiration for media bias detection. 
Still, to overcome the abovementioned drawbacks, we need to adapt the strategies from existing benchmarks, which we describe in \Cref{sec:dataset-selection}.
We base our \mbib guidelines on the best-practice benchmark examples, mainly on SuperGLUE \cite{wangSuperGLUEStickierBenchmark2019}: Every task should have a well-defined metric, and all tasks should have public training data available.
Even more, the tasks should be too difficult to solve for current solutions but should be solvable by humans.\footnote{This can sometimes also mean that only trained humans can solve the task.}
Finally, tasks should be in a format that is as simple as possible.

\section{Building MBIB}\label{sec:buildingMbib} 
In the following, we detail the process that leads to the creation of \mbib. 
First, we collect and select relevant media bias tasks and related datasets (\Cref{sec:media bias tasks}). 
The tasks should cover media bias as comprehensively as possible while reflecting societal relevance and existing research priorities. 
For instance, gender bias in the news has become a focus of bias research \cite{vanderpas2020}.
Second, we refine the initial list of datasets through a detailed study (\Cref{sec:dataset collection}). 
Mainly, we incorporate factors such as dataset availability, size, and quality. 
The avoidance of duplicates (using the same data basis or the same data set) is also taken into account here.
Third, we preprocess and unify the chosen datasets (\Cref{sec:dataset-preprocessing}). 
We also give a detailed overview of the properties of the selected datasets (\Cref{sec:dataset Properties}).
Finally, we conclude the construction of \mbib with a framework defining how models can be evaluated based on the tasks (\Cref{sec:evaluation and baselines}). 
We set a transformer-based baseline performance within the framework on all \mbib tasks.

\subsection{Media Bias Tasks}
\label{sec:media bias tasks}
A comprehensive curation of tasks that address media bias is challenging as it encompasses various forms of bias.
To ensure a systematic examination, each task should tackle an independent subcategory of media bias detection (e.g., linguistic or gender bias).
Specifying media bias types into categories facilitates each task's definition, delimitation, and interpretation. 
We select tasks based on two criteria:

\begin{enumerate}
     \item When occurring in media coverage, the task's associated bias should constitute a form of media bias.
     \item Tasks should either (A) be part of the media bias framework presented by \citet{spindeIntroducingMediaBias2022}, or (B) be an independent, distinguishable research field of societal importance.\footnote{This criterion implies that the tasks are not necessarily mutually exclusive.}
\end{enumerate}

To identify task candidates,  we assess a list of 322 media bias-related publications mentioned in  \cite{spindeIntroducingMediaBias2022}.\footnote{The authors automatically and then manually filter all relevant media bias research published between 2018 and 2023 \cite{spindeIntroducingMediaBias2022}. In total, they consider over 100,000 publications.}  
To the best of our knowledge, it is the most extensive existing literature collection on media bias to date \cite{spindeIntroducingMediaBias2022}. 
Their list categorizes the publications by the type of bias they focus on. 
We use these (task) categories as a starting point for our task selection, given that they all fulfill criterion (1). 
Also, all tasks fulfill either criterion (2A) or (2B).
In total, we identify the following task candidates: linguistic, text-level context, reporting-level context, cognitive bias (by criterion (2A)), as well as hate speech, fake news, political bias, racial bias, gender bias, religious bias, group bias, and framing effects (by criterion (2B)).

Of the task candidates mentioned above, framing effects, group bias, and religious bias are not included in the \mbib task. 
Group bias is an umbrella term describing bias introduced when dealing with various groups and includes gender bias, racial bias, and religious bias.
Instead of group bias, we, therefore, include the single tasks to avoid repetitions. 
Framing effects refer to how media organizations present information to influence the reader's perception.
We do not include framing effects due to their similarity to framing bias being a component of the linguistic bias task.
We decided that religious bias is not part of \mbib due to the low research interest (only one associated paper).
Also, we do not include a wide range of media bias-related tasks, such as sentiment analysis or stance detection \cite{spinde-exploiting-2021} since they do not directly identify forms of media bias. 
Especially sentiment detection is already covered in multiple existing benchmarks \cite{potts-etal-2021-dynasent}. 
The remaining five task candidates fulfilling criterion (2B) are added as external tasks to \mbib, together with the four tasks fulfilling criterion (2A). Therefore, \mbib consists of 9 tasks in total. 

In the remainder of this section, we briefly introduce the chosen tasks. More details and respective subcategories of each task can be found in \cite{spindeIntroducingMediaBias2022}.
For every task, \Cref{tab:biases} shows an example.\\

\noindent
\textbf{Linguistic bias} encompasses all forms of bias induced by lexical features, such as word choice and sentence structure, often subconsciously used \cite{spindeIntroducingMediaBias2022}. 
Generally, linguistic bias is expressed through specific word choice that reflects the social-category cognition applied to any described group or individual(s) \cite{beukeboom_linguistic_2017}. 

\noindent 
\textbf{Text-level context bias} refers to the expression of a text's context, whereby words and statements can shape the context of an article and sway the reader's perspective \cite{hubeNeuralBasedStatement2019}. 
These biases can be used to portray a particular opinion in a biased way by criticizing one side more than the other, using inflammatory words, or omitting relevant information. 

\noindent 
\textbf{Reporting-level context bias} refers to bias that arises through decisions made by editors and journalists on what events to report and which sources to use \cite{dalessioMediaBiasPresidential2000a}.
While text-level context bias examines the bias present within an individual article, reporting-level bias focuses on systematic attention given to specific topics. 

\noindent 
\textbf{Cognitive bias} occurs when readers introduce bias by selecting which articles to read and which sources to trust, which can be amplified in social media \cite{nickersonConfirmationBiasUbiquitous1998}.
These biases can lead to self-reinforcing cycles and expose readers to only one side of an issue.

\noindent 
\textbf{Hate speech} refers to any language that manifests hatred towards a specific group or aims to degrade, humiliate, or offend \cite{mathewHateXplainBenchmarkDataset2021a}.
Usually, hate speech is induced by using linguistic bias \cite{mozafari2020hate}.
Particularly in social media, the impact of hate speech is significant and exacerbates tensions between involved parties. 
However, similar processes can also be observed within, e.g., comments on news websites \cite{zannettouMeasuringCharacterizingHate2020}.

\noindent 
\textbf{Fake news} refers to published content based on false claims and premises, presented as being true to deceive the reader \cite{tandocjr.FactsFakeNews2019}. 
Research on fake news detection typically focuses on detecting it through linguistic features or comparing content to verified information \cite{tandocjr.FactsFakeNews2019}. 
Fake news have serious consequences, such as potential influences on the readers' health and political decisions \cite{rochaImpactFakeNews2021}.
However, the exact overlap between media bias and fake news is yet unclear; we address this again in \Cref{sec:discussion}.

\noindent 
\textbf{Racial bias} is expressed through negative or positive portrayals of racial groups. 
Research has shown that racial bias in news coverage can severely impact affected minorities, such as strengthening stereotypes and discrimination \cite{fairWarFaminePoverty1993, minMissingChildrenNational2010}. 

\noindent 
\textbf{Gender bias} in media can manifest as discrimination against one gender through underrepresentation or negative portrayal. 
Gender bias in media can severely impact perceptions of professions and role models \cite{singhFemaleLibrariansMale2020a}, as well as voting decisions \cite{laveryGenderBiasMedia2013}.

\noindent 
\textbf{Political bias} refers to a text's political leaning or ideology, potentially influencing the reader's political opinion and, ultimately, their voting behavior \cite{feldmanPoliticalIdeology2013}. 
There are several approaches to detecting political bias in media, e.g., counting the appearance of certain political parties or ideology-associated words.

\begin{table}[ht]
\centering
\caption{Media bias tasks and examples}
\begin{tabularx}{\linewidth}{@{}m{0.34\linewidth}@{}m{0.65\linewidth}}
\toprule
Task & Example from the \mbib datasets\\
\midrule
Linguistic bias &``A Trump-loving white security guard with a racist past shot and killed an unarmed Black man during an unprovoked hotel parking lot attack.'' \cite{spindeNeuralMediaBias2021} \\
\makecell[l]{Text-level Context\\Bias} & ``The governor [...] observed an influx of Ukrainian citizens who want to stay in Russia until the situation normalises in their country'' \cite{farberMultidimensionalDatasetBased2020a} \\
\makecell[l]{Reporting-Level\\Context Bias} & In a presidential campaign, one candidate receives disproportionate news coverage. \cite{dalessioMediaBiasPresidential2000a}\\
Cognitive Bias & ``Republicans are certain that the more people learn the less they'll like about the Democrats approach'' \cite{liuPOLITICSPretrainingSamestory2022} \\
Hate Speech & ``I will call my friends and we go [vulgarity] up that [vulgarity]'' \cite{mathewHateXplainBenchmarkDataset2021a} \\
Racial Bias & ``black people have a high crime rate therefore black people are criminals'' \cite{barikeriRedditBiasRealWorldResource2021a} \\
Fake News &``Phoenix Arizona is the No 2 kidnapping capital of the world'' \cite{wangLiarLiarPants2017a} \\
Gender Bias &``For a woman that is good.'' \cite{groszAutomaticDetectionSexist2020} \\
Political Bias & ``Generally happy with her fiscally prudent, dont-buy-what-you-cant-afford approach [...]'' (classified right) vs ``[...] some German voters have also begun to question austerity.'' (classified left) \cite{liuPOLITICSPretrainingSamestory2022} \\
\bottomrule
\end{tabularx}
\label{tab:biases}
\end{table}

\subsection{Dataset Collection}
\label{sec:dataset collection}

Based on our tasks, we select suitable data for each task. 
These should be available and widely used datasets in line with the guideline by \citet{wangSuperGLUEStickierBenchmark2019}. 
Furthermore, datasets should be diverse in the type of bias they cover and have consistent, high-quality annotations.
Since no extensive overview of the datasets used in the domain exists to date, we once again assess the list of media bias-related publications mentioned in the literature review by \citet{spindeIntroducingMediaBias2022} previously used for the task selection. 

In our work, we manually analyze all articles in the list to assess each used dataset, providing a complete overview of utilized datasets.
We review 322 media bias-related articles, resulting in a dataset overview of 115 datasets.

\begin{figure}[ht!]
\caption{The dataset collection and selection process}
\makebox[\textwidth][l]{\includegraphics[width=\linewidth]{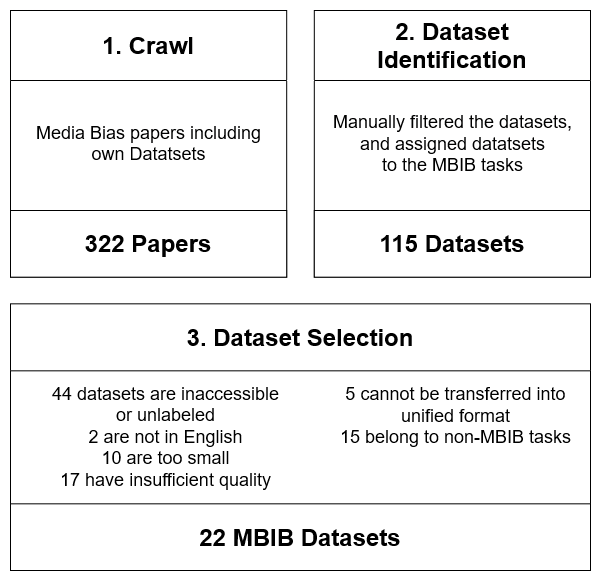}}
\label{img:dataset process}
\end{figure}

\subsection{Dataset Selection}
\label{sec:dataset-selection}
In benchmarks like SuperGLUE \cite{wangSuperGLUEStickierBenchmark2019}, and BigBench \cite{srivastavaImitationGameQuantifying2022}, one dataset is used per task. 
Having only a single dataset requires a dataset for every task that is sufficient in size and quality, covering the entire task. 
For media bias, as described in \Cref{sec:benchmarks}, no such single dataset per task exists.
Either datasets are too small, or they only cover a particular aspect of a task (e.g., datasets associated to the linguistic bias task either cover framing bias or connotation bias but not both). 
Therefore, we base every task on multiple datasets containing varying definitions of bias, reflecting the dataset-scarce research area as detailed as possible. 
The variety of initial datasets, which we detail more in \Cref{sec:dataset Properties}, gives more reason why a benchmark and overview of the domain is required.   
Mainly, we evaluate our datasets based on the following criteria: 

\begin{enumerate}
    \item First, we identify whether a dataset is accessible\footnote{We consider a dataset accessible if it is either publicly available, can be directly recreated (e.g., tweet IDs and labels provided), or there is a defined way to obtain it from the authors.} and labeled.
    \item Second, we identify whether the dataset uses the English language. For now, we focus only on English since it is the dominant language in the research domain \cite{SpindeGerman2020}. 
    \item Third, we evaluate the dataset size.  
    While bigger datasets usually contain a more balanced range of content, they are often labeled automatically. 
    Smaller datasets are mostly manually labeled and of higher quality. 
    However, they do not exhibit sufficient data points for many current system architectures \cite{spindeNeuralMediaBias2021}. We set a minimum of no less than 700 data points per dataset\footnote{After manual inspection, we conclude that the RacialBias \cite{ghoshalRacialBiasTwitter2018} dataset is the smallest set that still included high-quality annotations with sufficient variety in content.}. 
    \item Fourth, we manually evaluate the dataset quality, in terms of dataset transparency, diversity of sources, overlap with other datasets, and potential annotator training. We summarize the benefits of all chosen datasets within the \hyperref[mbiburl]{MBIB repository}. We are aware that ultimately, the assessment of quality in \mbib is based on a manual choice, and address this again in \Cref{sec:discussion}.
    \item Fifth, we require that datasets can be transformed into one unified format as specified in the remainder of this section.
    \item Sixth, datasets need to belong to one of the \mbib's tasks.
\end{enumerate}

\Cref{img:dataset process} includes the number of datasets filtered out based on these criteria. 
Of the 115 datasets collected, only 74 (65\%) are directly available. 
We discard all 41 others, as well as three which have no labels.
Two datasets are not in English.
We discard five datasets because they could not be transformed into a unified format.\footnote{For instance, the dataset provided by \cite{breitfeller-etal-2019-finding} contains quotes with associated statements outlining the context, which could not be transformed into a binary label.}
Finally, 15 datasets belong to non-\mbib tasks such as sentiment analysis.
After applying the criteria, 22 datasets remain.
\Cref{tab:categories and datasets} displays the selected datasets for each task with their respective size.
Out of the 22 selected datasets two can only be obtained directly from the authors due to copyright issues \cite{golbeckLargeLabeledCorpus2017, liuPOLITICSPretrainingSamestory2022}.
\mbib is, therefore, currently available in two versions: in a base and a full module. 
The base version includes only the datasets directly available and aims to facilitate access.
In the full version all 22 datasets are used. 
We provide guidance on accessing and preprocessing the remaining datasets to create the full version in the \hyperref[mbiburl]{MBIB repository}. 
All our experiments (\autoref{sec:benchmarks}) use the full version of MBIB.


After the dataset selection, no datasets associated with reporting-level context bias are left. 
Of the initial six datasets for reporting-level context bias, five are not publicly available, and the only available dataset is not labeled \cite{Hofmann2021}. 
Reporting-level context bias, therefore, is currently not included in \mbib.
We aim to add reporting-level context bias to \mbib as soon as enough data exists.

\begin{table}[h]
\caption{\small{Tasks and datasets in \mbib}}\centering
\small
\begin{tabular}{ll}
\toprule[1.5pt]
 Tasks and Datasets	&	Data Points	\\
 \midrule
 \textbf{Linguistic Bias}	& \textbf{433,677*} \tnote{*}\\
\hspace{5mm}Wikipedia NPOV \cite{hubeNeuralBasedStatement2019b}	&	11,945	\\
\hspace{5mm}BABE \cite{spindeNeuralMediaBias2021}	&	3,673	\\
\hspace{5mm}Wiki Neutrality Corpus \cite{pryzantAutomaticallyNeutralizingSubjective2019}	&	362,991	\\
\hspace{5mm}UsVsThem \cite{huguetcabotUsVsThem2021}	&	6,863	\\
\hspace{5mm}RedditBias \cite{barikeriRedditBiasRealWorldResource2021a}	&	10,583	\\
\hspace{5mm}Media Frames Corpus \cite{kwakSystematicMediaFrame2020}	&	37,622	\\
\hspace{5mm}BASIL \cite{fanPlainSightMedia2019}	&	1,726	\\
\hspace{5mm}Biased Sentences \cite{zotero-1084}	&	842	\\

\textbf{Cognitive Bias}	&	\textbf{2,344,387*} \tnote{*}	\\
\hspace{5mm}BIGNEWS \cite{liuPOLITICSPretrainingSamestory2022}	&	2,331,552	\\
\hspace{5mm}Liar Dataset \cite{wangLiarLiarPants2017a}	&	12,835	\\

\textbf{Text-Level Context Bias}	&	\textbf{28,329*} \tnote{*}	\\
\hspace{5mm}Contextual Abuse Dataset \cite{vidgenIntroducingCADContextual2021}	&	26,235	\\
\hspace{5mm}Multidimensional Dataset \cite{farberMultidimensionalDatasetBased2020a}	&	2,094	\\

\textbf{Hate Speech}	&	\textbf{2,050,674*} \tnote{*}	\\
\hspace{5mm}Kaggle Jigsaw \cite{jigsaw/conversationaiJigsawUnintendedBias2019}	&	1,999,516	\\
\hspace{5mm}HateXplain \cite{mathewHateXplainBenchmarkDataset2021a}	&	20,148	\\
\hspace{5mm}RedditBias \cite{barikeriRedditBiasRealWorldResource2021a}	&	10,583	\\
\hspace{5mm}Online Harassment Corpus \cite{golbeckLargeLabeledCorpus2017}	&	20,427	\\

\textbf{Gender Bias}	& \textbf{33,121*} \tnote{*}	\\
\hspace{5mm}RedditBias \cite{barikeriRedditBiasRealWorldResource2021a}	&	3,000	\\
\hspace{5mm}RtGender \cite{voigtRtGenderCorpusStudying2018}	&	15,351	\\
\hspace{5mm}WorkPlace sexism \cite{groszAutomaticDetectionSexist2020}	&	1,136	\\
\hspace{5mm}CMSB \cite{samoryCallMeSexist2020}	&	13,634	\\

\textbf{Racial Bias}	& \textbf{2,371*} \tnote{*}	\\
\hspace{5mm}RedditBias \cite{barikeriRedditBiasRealWorldResource2021a}	&	2,620	\\
\hspace{5mm}RacialBias \cite{ghoshalRacialBiasTwitter2018}	&751	\\

\textbf{Fake News}&\textbf{24,394*} \tnote{*}	\\
\hspace{5mm}Liar Dataset \cite{wangLiarLiarPants2017a}	&	12,835	\\
\hspace{5mm}PHEME \cite{zubiagaExploitingContextRumour2017}	&	5,222	\\
\hspace{5mm}FakeNewsNet \cite{shuFakeNewsNetDataRepository2020}	&	6,337	\\

\textbf{Political Bias}&\textbf{2,348,198*} \tnote{*}	\\
\hspace{5mm}UsVsThem \cite{huguetcabotUsVsThem2021}	&	6,863	\\
\hspace{5mm}BIGNEWS \cite{liuPOLITICSPretrainingSamestory2022}	&	2,331,552	\\
\hspace{5mm}SemEval \cite{kieselDataPANSemEval2018}	&	9,783	\\

\bottomrule[0.5pt]
\end{tabular}
\begin{tablenotes}
\item[*] *Refers to the total number of data points of the task.
\end{tablenotes}
\label{tab:categories and datasets}
\end{table}

\subsection{Preprocessing}
\label{sec:dataset-preprocessing}
We preprocess all of the above datasets into one unified format consisting of the unique ID of the text segment to be analyzed, an ID indicating to which dataset the statement belongs, the text, a binary label, and, if given, additional labels\footnote{We show the exact preprocessing steps for every dataset within the \hyperref[mbiburl]{MBIB repository}.
There, we also show the datasets in more detail.}. 

While keeping the original labels, we transform all dataset labels into a binary label format.
This has three major advantages:
\begin{enumerate}
     \item It allows an easy combination of different datasets without requiring different model heads. 
     \item It follows the task principles set up by \citet{wangSuperGLUEStickierBenchmark2019} to formulate the task as simple as possible.
     \item By keeping the original labels, changes applied to non-binary labels can be tracked transparently\footnote{Investigating the non-binary labels in more detail will also be an interesting aspect in future work.}. 
\end{enumerate}

Most datasets already have binary labels. 
However, we determine a threshold for some datasets with continuous labels to binarize the data. 
If possible, the authors' recommendation for a threshold is followed. 
The original format of every dataset is shown in \autoref{tab:datasets details}.
For instance, the Kaggle Jigsaw data \cite{jigsaw/conversationaiJigsawUnintendedBias2019} is labeled on a scale from 0 to 1. 
The authors recommend using a 0.5 threshold to binarize the label, which we follow for \mbib. 
Also, we collapse multi-categorical labels into two categories. 
For instance, for the political bias task, `right' and `left' are combined into `biased' \cite{liuPOLITICSPretrainingSamestory2022, kieselDataPANSemEval2018}. 
The Liar dataset \cite{wangLiarLiarPants2017a} provides even more labels: `true', `mostly-true', `half-true', `barely-true', `false', and `pants-fire'.
The first four labels are combined into a single `true' label and the last two into one `false' label. 
Even though social media-specific elements such as hashtags or emoticons are used in related areas such as sentiment analysis \cite{wankhede-2018}, we remove them to not deviate further from a news format. 
As an additional step, we enrich the FakeNewsNet dataset by scraping the tweets or articles referred to by the Tweet IDs given in the original resource \citep{shuFakeNewsNetDataRepository2020}\footnote{The original dataset does only include IDs, not the texts themselves.}. 
Every decision with regards to the unified format is furthermore detailed in the \hyperref[mbiburl]{MBIB repository}.

\section{Dataset Properties}
\label{sec:dataset Properties}
Out of the 115 datasets in the dataset overview, 38 contain annotated articles, 32 annotated sentences, 25 annotated social media posts (mainly Tweets), five annotated comments, and two annotated headlines. 
The most prominent article source for datasets is \url{allsides.com} (eight datasets). 
Five datasets stem from Wikipedia. 
Ten datasets are created using Amazon Mechanical Turk workers; six refer to other crowd-working platforms. 
32 datasets are found to be either self-annotated by the authors or by individually hired annotators. 
20 datasets are labeled using distant labeling (retrieving the label for a single article from the outlet's bias score). 
Overall, we find binary, multi-class, and continuous labels. 
The data also contains a lot of other annotations, such as bias-inducing words or context data. 
The datasets have a median of 8,656 data points, with significant variance.
The dataset distribution among \mbib's tasks can be found in \Cref{img:dataset_distr}. 

\begin{figure}[ht!]
\caption{Dataset distribution over \mbib tasks}
\makebox[\textwidth][l]{\includegraphics[width=\linewidth]{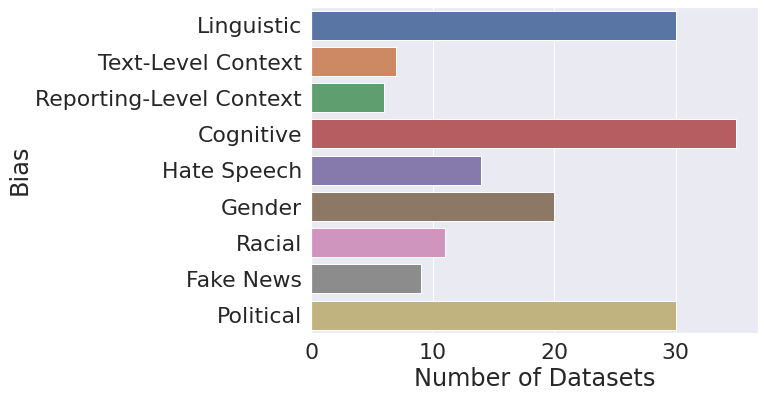}}
\end{figure}
\label{img:dataset_distr}

Out of the 22 datasets in \mbib, nine contain news articles.
Eleven include data from social media, two data from Wikipedia, and one dataset only consists of quotes (one dataset has two different sources).
More details on the properties of all datasets can be found in \autoref{tab:datasets details}.
The social media datasets use data from Reddit \cite{huguetcabotUsVsThem2021, barikeriRedditBiasRealWorldResource2021a, vidgenIntroducingCADContextual2021, voigtRtGenderCorpusStudying2018, groszAutomaticDetectionSexist2020} and Twitter \cite{jigsaw/conversationaiJigsawUnintendedBias2019, golbeckLargeLabeledCorpus2017, mathewHateXplainBenchmarkDataset2021a, samoryCallMeSexist2020, ghoshalRacialBiasTwitter2018, zubiagaExploitingContextRumour2017}. 
Only \cite{voigtRtGenderCorpusStudying2018, mathewHateXplainBenchmarkDataset2021a} further include data from other platforms such as Facebook and Gab\footnote{\url{https://gab.com} is a microblogging and social networking service.}. 
Also, the social media datasets usually focus on specific events \cite{huguetcabotUsVsThem2021}, or phrases \cite{mathewHateXplainBenchmarkDataset2021a, samoryCallMeSexist2020}. 
The Wikipedia-based datasets \cite{hubeNeuralBasedStatement2019b, pryzantAutomaticallyNeutralizingSubjective2019} are both based on Wikipedia's POV label, signaling a potentially biased statement.  
News articles within the collection come from various widely known sources such as the New York Times \cite{liuPOLITICSPretrainingSamestory2022} and alternative media sources \cite{wangLiarLiarPants2017a}. 
Partially the datasets are general collections of news articles \cite{kwakSystematicMediaFrame2020, fanPlainSightMedia2019, wangLiarLiarPants2017a, liuPOLITICSPretrainingSamestory2022, shuFakeNewsNetDataRepository2020, kieselDataPANSemEval2018} and partially they contain articles collected around certain topics \cite{farberMultidimensionalDatasetBased2020a, spindeNeuralMediaBias2021, zotero-1084}. 

For the annotations, most authors use crowdsourcing \cite{hubeNeuralBasedStatement2019b, huguetcabotUsVsThem2021, kwakSystematicMediaFrame2020, vidgenIntroducingCADContextual2021, farberMultidimensionalDatasetBased2020a, mathewHateXplainBenchmarkDataset2021a, spinde-tassy-2021, voigtRtGenderCorpusStudying2018, samoryCallMeSexist2020}.
Others train or commission selected annotators \cite{spindeNeuralMediaBias2021, kieselDataPANSemEval2018}, annotate themselves \cite{groszAutomaticDetectionSexist2020, ghoshalRacialBiasTwitter2018} or use external annotations \cite{pryzantAutomaticallyNeutralizingSubjective2019, wangLiarLiarPants2017a, liuPOLITICSPretrainingSamestory2022, shuFakeNewsNetDataRepository2020}. 
Some contributions further report either an annotator training, instructions, or mechanisms to ensure high annotation quality (such as control questions) \cite{spindeNeuralMediaBias2021, hubeNeuralBasedStatement2019b, barikeriRedditBiasRealWorldResource2021a, fanPlainSightMedia2019, farberMultidimensionalDatasetBased2020a, mathewHateXplainBenchmarkDataset2021a, golbeckLargeLabeledCorpus2017}. 
The labels provided range from binary labels \cite{hubeNeuralBasedStatement2019, fanPlainSightMedia2019, spindeNeuralMediaBias2021, zubiagaExploitingContextRumour2017, shuFakeNewsNetDataRepository2020, pryzantAutomaticallyNeutralizingSubjective2019, barikeriRedditBiasRealWorldResource2021a, vidgenIntroducingCADContextual2021, golbeckLargeLabeledCorpus2017, mathewHateXplainBenchmarkDataset2021a, groszAutomaticDetectionSexist2020, ghoshalRacialBiasTwitter2018} to multi-class labels  \cite{farberMultidimensionalDatasetBased2020a, kieselDataPANSemEval2018, kwakSystematicMediaFrame2020, zotero-1084, liuPOLITICSPretrainingSamestory2022, wangLiarLiarPants2017a, voigtRtGenderCorpusStudying2018} and continuous labels \cite{jigsaw/conversationaiJigsawUnintendedBias2019, huguetcabotUsVsThem2021, samoryCallMeSexist2020}. \\
The BigNews Corpus \cite{liuPOLITICSPretrainingSamestory2022}, the largest dataset in our collection, is the only dataset not directly labeled on the individual text level. 
Instead, annotations are based on the outlet's bias label. 

\begin{table*}[t]
\centering
\caption{Details of \mbib's datasets}
\begin{tabularx}{\textwidth}{lllll}
\toprule
Dataset Name  & Feature Level & Feature Source & Label Categories & Label Source \\
\midrule
Wikipedia NPOV \cite{hubeNeuralBasedStatement2019b}  & statements & Wikipedia & binary bias label & Wikipedia editors \\
BASIL \cite{fanPlainSightMedia2019} & event spans & news articles & binary bias label & trained annotators \\
BABE \cite{spindeNeuralMediaBias2021} & sentences & news articles & binary bias label & trained annotators \\
PHEME \cite{zubiagaExploitingContextRumour2017} & tweets & Twitter & \makecell[l]{binary rumor and\\veracity label} & journalists \\
Multidimensional Dataset \cite{farberMultidimensionalDatasetBased2020a} & sentences & news articles& \makecell[l]{labeled on three\\bias dimensions} & crowdsourcing, expert control \\
FakeNewsNet \cite{shuFakeNewsNetDataRepository2020} & sentences & news articles& binary veracity label & fact-checking websites \\
Wiki Neutrality Corpus \cite{pryzantAutomaticallyNeutralizingSubjective2019} & sentences & Wikipedia & binary bias label & Wikipedia editors \\
SemEval \cite{kieselDataPANSemEval2018} & sentences & news articles & \makecell[l]{multi-categorical\\hyperpartisan label} & three annotators \\
Media Frames Corpus \cite{kwakSystematicMediaFrame2020} & sentences & news articles & neutral/pro/anti & crowdsourcing \\
Biased Sentences Dataset \cite{zotero-1084}	 & sentences & news articles & \makecell[l]{multi-categorical\\bias label} & crowdsourcing \\
Kaggle Jigsaw \cite{jigsaw/conversationaiJigsawUnintendedBias2019}& comments & Twitter & continuous toxicity label & annotators \\
UsVsThem \cite{huguetcabotUsVsThem2021} & comments & Reddit & continuous bias label & crowdsourcing\\
BIGNEWS \cite{liuPOLITICSPretrainingSamestory2022}	 & sentences & news articles & left/neutral/right & allsides.com \\
Liar Dataset \cite{wangLiarLiarPants2017a} & statements & politifacts.com & \makecell[l]{multi-categorical\\veracity label} & expert annotators \\
RedditBias \cite{barikeriRedditBiasRealWorldResource2021a} & comments & Reddit & binary bias label & trained annotators \\
Contextual Abuse Dataset \cite{vidgenIntroducingCADContextual2021} & posts, comments & Reddit & binary abuse label & expert annotators \\
Online Harassment Corpus \cite{golbeckLargeLabeledCorpus2017}& tweets & Twitter & binary harassment label & trained annotators \\
HateXplain \cite{mathewHateXplainBenchmarkDataset2021a} & sentences & Twitter and Gab & binary hatespeech label &crowdsourcing\\
RtGender \cite{voigtRtGenderCorpusStudying2018} & sentences & \makecell[l]{Facebook, Reddit,\\TED, Fitocracy} & \makecell[l]{multi-categorical gender\\perception label} & crowdsourcing \\
WorkPlace sexism \cite{groszAutomaticDetectionSexist2020}& sentences & news articles, quotes & binary gender bias label & trained annotators \\
CMSB \cite{samoryCallMeSexist2020}& tweets & Twitter & continuous sexism scales &crowdsourcing\\
RacialBias \cite{ghoshalRacialBiasTwitter2018}& tweets & Twitter & binary racial bias label & annotators \\
\bottomrule[0.5pt]
\end{tabularx}

\label{tab:datasets details}
\end{table*}

\section{Evaluation and Baselines}
\label{sec:evaluation and baselines}
\subsection{Evaluation Framework}
To evaluate models on \mbib, we introduce a framework defining which metrics should be used and reported.
We illustrate the usage of our framework by evaluating five transformer models on \mbib.  
We use stratified 5-fold-cross-validation on the preprocessed \mbib data, which ensures stable scores while remaining computationally feasible for larger tasks.
Also, we balance the classes in each task to ensure an equal representation of classes in each fold and thus ensure unbiased scores.

As the primary performance metric, we choose the $F_{1}$-score based on its established usage as a metric in various benchmarks, including those previously discussed \cite{wangSuperGLUEStickierBenchmark2019, wangGLUEMultiTaskBenchmark2019, srivastavaImitationGameQuantifying2022}. 
All scores of the five folds are averaged.
As the datasets we combine into one task differ in the number of observations they contain, larger datasets can strongly influence the final score.
Therefore, to calculate the $F_{1}$-scores, we propose two methods:
\begin{itemize}
    \item The micro average $F_{1}$-score: One $F_{1}$-score is calculated on the predictions of a model on the entire test set. 
    The scores of the five folds are averaged.
    \item The macro average $F_{1}$-score: Multiple $F_{1}$-scores are calculated on the predictions of a model on the test set. 
    One $F_{1}$-score is calculated individually for every dataset (from which the data originally stems). 
\end{itemize}
\noindent

The macro approach ensures that each dataset is represented equally in the final score, regardless of its size.\footnote{Which is particularly important due to the immense size differences between datasets, which likely skew the macro average $F_{1}$-scores towards our more extensive datasets.} 
The micro score's simplicity and focus on larger datasets enable an assessment of the impact of dataset size on the model's overall performance when used in conjunction with the macro score.
As performed in SuperGLUE \cite{wangSuperGLUEStickierBenchmark2019}, we average both scores for all tasks into two (micro and macro) final media bias scores.
By reporting the averaged scores, we can evaluate model generalizability. 
By also providing single-task scores, we can evaluate performances on individual tasks. 

\subsection{Testing Baselines}
\begin{figure}[ht!]
\centering
\caption{$F_{1}$-scores per task}
\begin{subfigure}[t]{0.45\textwidth}
\caption{Micro-average}
\centering
\includegraphics[width=\textwidth]{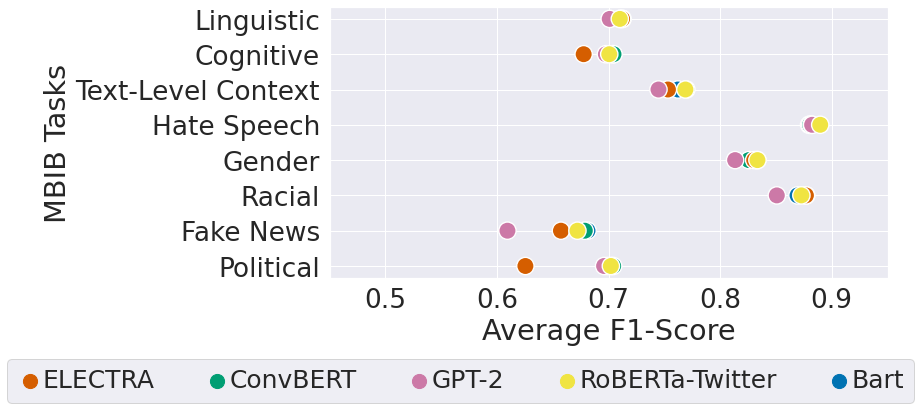}
\label{fig:micro_performance}
\end{subfigure}
\hfill
\begin{subfigure}[t]{0.45\textwidth}
\centering
\caption{Macro-average}
\includegraphics[width=\textwidth]{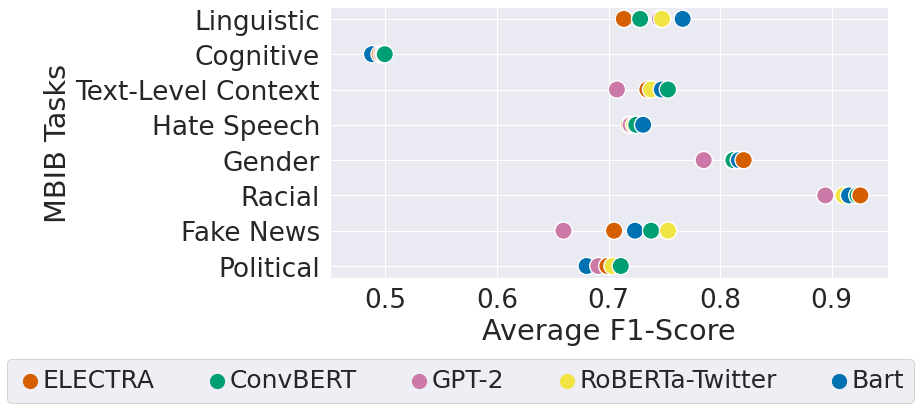}
\label{fig:macro_performance}
\end{subfigure}
\label{fig:performance}
\end{figure}

As baselines, we test base versions of five available transformer models, ConvBERT, Bart, RoBERTa-Twitter, ELECTRA, and GPT-2, focusing only on lexical features. \footnote{So, for instance, for fake news detection, we do not use fact-checking databases. We provide details on these models and why we choose them in the \hyperref[mbiburl]{MBIB repository}} 
\autoref{fig:performance} displays the performance of the five evaluated models on every task.
The results of the best performing models can furthermore be found in \Cref{tab:BaselinePerformances}.\footnote{We show the average final scores within the \hyperref[mbiburl]{MBIB repository}.}  
The baseline results give a first intuition about the purpose of \mbib as they show that no single transformer model stands out as the best-performing model across all tasks.
We find substantial inter-task performance differences.
These differences indicate that some tasks, e.g., racial and gender bias, seem easier to detect than other tasks, such as fake news or cognitive bias.
However, the performances are very similar within individual tasks.
GPT-2 underperforms on most tasks.
The distinction between the micro and macro F1-scores highlights a more comprehensive evaluation of a model's ability to detect media bias, revealing its specific strengths and weaknesses. However, it is crucial to explore alternative metrics and delve deeper into the analysis of individual task scores, as further research in this area remains necessary.
When creating \mbib, a key concern was that the varying size of the datasets in a task could disproportionately affect the performance of the models.
Therefore, the performance on individual datasets is considered in relation to the size of the datasets (\Cref{img:size_vs_performance}).
A positive linear relationship could be expected if there were a direct correlation between size and performance.  
However, this cannot be observed.\\
We foresee advancements in the performance on the \mbib and encourage continuous assessment of innovative and refined techniques to maintain this progress.

\section{Discussion}\label{sec:discussion}
Already in 2021, \citet{spindeNeuralMediaBias2021} emphasized that media bias systems with better generalization capabilities are needed. 
The authors concluded that a way to promote such generalization capabilities would be a more refined evaluation scheme, as presented in CheckList \citep{ribeiro-etal-2020-beyond}. 
\mbib, for the first time in the media bias domain, provides such a refined evaluation scheme and promotes the development of models with stronger generalizability.
Also, \mbib illustrates that many different tasks exist in the media bias domain, and creates awareness of media bias being a complex construct in total.

While the media bias framework of \cite{spindeIntroducingMediaBias2022} helps to capture this complexity, it has the downside of tasks not being intuitive to understand, sometimes causing unclarity about where a specific bias fits in.
Future work should, therefore, improve taxonomies and conceptual overviews of the research domain.
The amalgamation of various bias tasks presents a considerable challenge due to factors such as legal disparities, regional variations, and cultural contexts. Addressing these complexities necessitates meticulous dataset curation, annotation guidelines, and a thoughtful approach to evaluation procedures. Nevertheless, it is through this comprehensive combination that we can foster the development of methodologies capable of addressing multiple types of biases concurrently.
For the selection of datasets, we aim to present a stringent justification.
However, other compositions of datasets are also conceivable. 
While some areas in the domain exhibit multiple datasets, others are scarce or have no particular datasets targeting the respective subconcept.
The lack of datasets for the reporting-level context bias task additionally limits \mbib. 
Also, the text-level context and racial bias tasks can be strengthened by more available datasets as they currently only contain two datasets.
Especially for text-level context this leaves the informational bias aspect (as described by \cite{spindeIntroducingMediaBias2022}) uncovered.
We hope that our comprehensive dataset overview can promote the creation of further datasets by facilitating the identification of areas with low data coverage. 
The large amount of social media data within our benchmark reflects that bias on social media is potentially stronger \cite{ghoshalRacialBiasTwitter2018}. 
However, to address social media and news outlets equally, both important sources of information, we propose to focus on developing more news article-related media bias datasets. 
Additionally, all datasets exhibit some tradeoff between manual labels being expensive, and automated labels not always being precise. 
Big datasets such as BIGNEWS \cite{liuPOLITICSPretrainingSamestory2022}, therefore, likely introduce a lot of noise due to less precise labels.
Generally, annotator training might help to increase agreement among the labels \cite{spindeNeuralMediaBias2021}, while quality control measures such as monitoring annotator performance might enhance overall data quality \cite{farberMultidimensionalDatasetBased2020a}. 
We acknowledge the importance of privacy and GDPR compliance in handling data from social media and online sources. 
All datasets included in MBIB are anonymized. 
To ensure continued compliance, we commit to periodically updating the benchmark to reflect changes in the source datasets, maintaining the integrity and relevance of our benchmark while respecting privacy regulations.
 
In future work on \mbib, other biases (like framing or religious bias) can be considered; even more, we envision the creation of a benchmark that also includes related concepts, such as sentiment, to continuously extend our benchmark to measure any opinion expressed in texts. 
A continued discussion about which tasks capture media bias best remains necessary to build such a future collection.  
Additionally, in future work, we will include multi-categorical or numerical data and integrate multiple languages into our benchmark \cite{spinde-integrated-2020}, which is yet only focusing on English. 
We want to increase the overall diversity of \mbib tasks so that it continues to address the complexity of media bias detection in more and more detail.

\section{Conclusion}\label{sec:conclusion}
This work proposes \mbib, the first ever multi-task benchmark for media bias. 
\mbib is organized over nine tasks and consists of 22 datasets, which we curate from a list of 115 datasets in total. 
We evaluate the datasets based on their focus, size, availability, and label quality to filter the original list. 
We also present a framework showing how future models can be evaluated using our benchmark. 
By introducing \mbib, we aim to capture media bias as extensively as possible and to give a complete overview of currently available resources in the domain. 
We believe that \mbib offers a new common ground for research in the domain, especially given the rising amount of (research) attention directed towards media bias.
We will continue to update \mbib in the future, and add potential new datasets and tasks.

\begin{acks}
Foremost, we want to thank Felix Kettenbeil, who helped us with the computing setup and revisions, as well as for valuable discussions about the project. We also thank Anna Bahß for proofreading in the final hours. The Hanns-Seidel-Foundation and the Friedrich-Ebert-Foundation, Germany, supported this work, as did the DAAD (German Academic Exchange Service). It was also supported by the Lower Saxony Ministry of Science and Culture and the VW Foundation. Furthermore, the authors would like to express their gratitude towards Prof. Isao Echizen and his research lab from the National Institute of Informatics in Tokyo, Japan, as well as to Prof. Juhi Kulshrestha for administrative support.
\end{acks}

\appendix

\section{Appendix}

\subsection{Model Baseline Results}

\begin{table}[H]
\caption{Best average scores per task}\centering
\begin{subtable}{\linewidth}
\caption{Micro-scores}\centering
\begin{tabular}{llc}
\toprule[1.5pt]
Bias Type & Model & Micro-Score \\
\midrule
Linguistic Bias & ConvBERT & 0.7126 \\
Cognitive Bias & ConvBERT & 0.7044 \\
Text-Level Context Bias & ConvBERT & 0.7697 \\
Hate Speech & RoBERTa-Twitter & 0.8897 \\
Gender Bias & RoBERTa-Twitter & 0.8334 \\
Racial Bias & ConvBERT & 0.8772 \\
Fake News & Bart & 0.6811 \\
Political Bias & ConvBERT & 0.7041 \\
\bottomrule[0.5pt]
\end{tabular}
\end{subtable}

\vspace{0.2cm}
\begin{subtable}{\linewidth}
\caption{Macro-scores}\centering
\begin{tabular}{llc}
\toprule[1.5pt]
Bias Type & Model & Macro-Score \\
\midrule
Linguistic Bias & Bart & 0.7664 \\
Cognitive Bias & ConvBERT & 0.4995 \\
Text-Level Context Bias & ConvBERT & 0.7532 \\
Hate Speech & Bart & 0.7310 \\
Gender Bias & ELECTRA & 0.8211 \\
Racial Bias & ELECTRA & 0.6170 \\
Fake News & RoBERTa-Twitter& 0.7533 \\
Political Bias & ConvBERT & 0.7110 \\
\bottomrule[0.5pt]
\end{tabular}
\end{subtable}
\label{tab:BaselinePerformances}
\end{table}

\subsection{Model Performance on the Dataset Level}

\begin{figure}[H]
\caption{$F_{1}$-scores per dataset and the size of the testset}
\makebox[\textwidth][l]{\includegraphics[width=\linewidth]{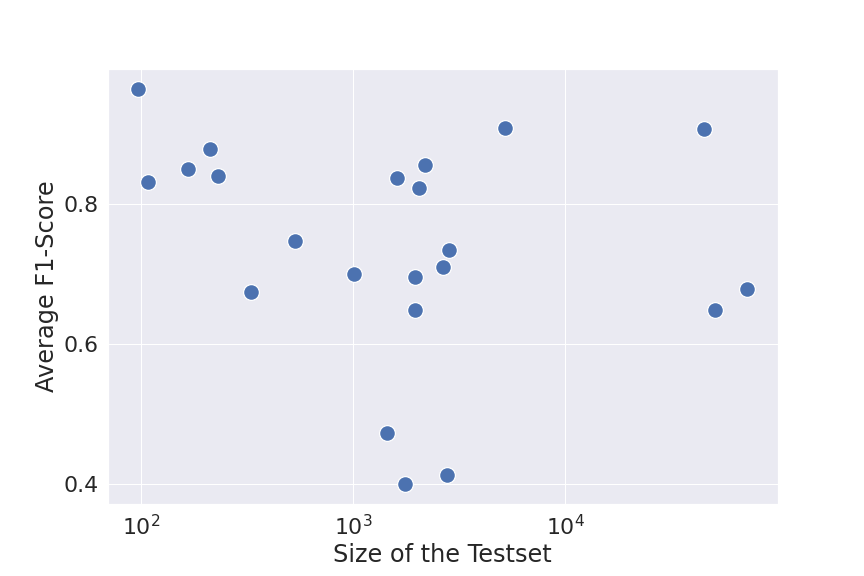}}
\end{figure}
\label{img:size_vs_performance}

\bibliographystyle{ACM-Reference-Format}
\balance
\bibliography{MBIB}


\begin{thebibliography}{60}


\ifx \showCODEN    \undefined \def \showCODEN     #1{\unskip}     \fi
\ifx \showDOI      \undefined \def \showDOI       #1{#1}\fi
\ifx \showISBNx    \undefined \def \showISBNx     #1{\unskip}     \fi
\ifx \showISBNxiii \undefined \def \showISBNxiii  #1{\unskip}     \fi
\ifx \showISSN     \undefined \def \showISSN      #1{\unskip}     \fi
\ifx \showLCCN     \undefined \def \showLCCN      #1{\unskip}     \fi
\ifx \shownote     \undefined \def \shownote      #1{#1}          \fi
\ifx \showarticletitle \undefined \def \showarticletitle #1{#1}   \fi
\ifx \showURL      \undefined \def \showURL       {\relax}        \fi
\providecommand\bibfield[2]{#2}
\providecommand\bibinfo[2]{#2}
\providecommand\natexlab[1]{#1}
\providecommand\showeprint[2][]{arXiv:#2}

\bibitem[AI(2019)]%
        {jigsaw/conversationaiJigsawUnintendedBias2019}
\bibfield{author}{\bibinfo{person}{Jigsaw/Conversation AI}.}
  \bibinfo{year}{2019}\natexlab{}.
\newblock \bibinfo{title}{Jigsaw unintended bias in toxicity classification.}
\newblock
\newblock
\urldef\tempurl%
\url{https://www.kaggle.com/competitions/jigsaw-unintended-bias-in-toxicity-classification/data}
\showURL{%
\tempurl}


\bibitem[Aribandi et~al\mbox{.}({[n.\,d.]})]%
        {aribandi_ext5_2021}
\bibfield{author}{\bibinfo{person}{Vamsi Aribandi}, \bibinfo{person}{Yi Tay},
  \bibinfo{person}{Tal Schuster}, \bibinfo{person}{Jinfeng Rao},
  \bibinfo{person}{Huaixiu~Steven Zheng}, \bibinfo{person}{Sanket~Vaibhav
  Mehta}, \bibinfo{person}{Honglei Zhuang}, \bibinfo{person}{Vinh~Q. Tran},
  \bibinfo{person}{Dara Bahri}, \bibinfo{person}{Jianmo Ni},
  \bibinfo{person}{Jai~Prakash Gupta}, \bibinfo{person}{Kai Hui},
  \bibinfo{person}{Sebastian Ruder}, {and} \bibinfo{person}{Donald Metzler}.}
  \bibinfo{year}{[n.\,d.]}\natexlab{}.
\newblock \showarticletitle{{ExT}5: Towards Extreme Multi-Task Scaling for
  Transfer Learning}.
\newblock   \bibinfo{volume}{abs/2111.10952} (\bibinfo{year}{[n.\,d.]}).
\newblock
\showeprint[arxiv]{2111.10952}
\urldef\tempurl%
\url{https://arxiv.org/abs/2111.10952}
\showURL{%
\tempurl}


\bibitem[Barikeri et~al\mbox{.}(2021)]%
        {barikeriRedditBiasRealWorldResource2021a}
\bibfield{author}{\bibinfo{person}{Soumya Barikeri}, \bibinfo{person}{Anne
  Lauscher}, \bibinfo{person}{Ivan Vulić}, {and} \bibinfo{person}{Goran
  Glavaš}.} \bibinfo{year}{2021}\natexlab{}.
\newblock \showarticletitle{{RedditBias}: {A} {Real}-{World} {Resource} for
  {Bias} {Evaluation} and {Debiasing} of {Conversational} {Language} {Models}}.
  In \bibinfo{booktitle}{\emph{Proceedings of the 59th {Annual} {Meeting} of
  the {Association} for {Computational} {Linguistics} and the 11th
  {International} {Joint} {Conference} on {Natural} {Language} {Processing}
  ({Volume} 1: {Long} {Papers})}}. \bibinfo{publisher}{Association for
  Computational Linguistics}, \bibinfo{address}{Online},
  \bibinfo{pages}{1941--1955}.
\newblock
\urldef\tempurl%
\url{https://doi.org/10.18653/v1/2021.acl-long.151}
\showDOI{\tempurl}


\bibitem[Baumer et~al\mbox{.}({[n.\,d.]})]%
        {baumer-testing-2015}
\bibfield{author}{\bibinfo{person}{Eric Baumer}, \bibinfo{person}{Elisha
  Elovic}, \bibinfo{person}{Ying Qin}, \bibinfo{person}{Francesca Polletta},
  {and} \bibinfo{person}{Geri Gay}.} \bibinfo{year}{[n.\,d.]}\natexlab{}.
\newblock \showarticletitle{Testing and Comparing Computational Approaches for
  Identifying the Language of Framing in Political News}. In
  \bibinfo{booktitle}{\emph{Proceedings of the 2015 Conference of the North
  American Chapter of the Association for Computational Linguistics: Human
  Language Technologies}} (Denver, Colorado, 2015).
  \bibinfo{publisher}{Association for Computational Linguistics},
  \bibinfo{pages}{1472--1482}.
\newblock
\urldef\tempurl%
\url{https://doi.org/10.3115/v1/N15-1171}
\showDOI{\tempurl}


\bibitem[Beukeboom and Burgers(2017)]%
        {beukeboom_linguistic_2017}
\bibfield{author}{\bibinfo{person}{Camiel~J. Beukeboom} {and}
  \bibinfo{person}{Christian Burgers}.} \bibinfo{year}{2017}\natexlab{}.
\newblock \showarticletitle{Linguistic {Bias}}.
\newblock In \bibinfo{booktitle}{\emph{Oxford {Research} {Encyclopedia} of
  {Communication}}}. \bibinfo{publisher}{Oxford University Press}.
\newblock
\showISBNx{978-0-19-022861-3}
\urldef\tempurl%
\url{https://doi.org/10.1093/acrefore/9780190228613.013.439}
\showDOI{\tempurl}


\bibitem[Breitfeller et~al\mbox{.}(2019)]%
        {breitfeller-etal-2019-finding}
\bibfield{author}{\bibinfo{person}{Luke Breitfeller}, \bibinfo{person}{Emily
  Ahn}, \bibinfo{person}{David Jurgens}, {and} \bibinfo{person}{Yulia
  Tsvetkov}.} \bibinfo{year}{2019}\natexlab{}.
\newblock \showarticletitle{Finding Microaggressions in the Wild: A Case for
  Locating Elusive Phenomena in Social Media Posts}. In
  \bibinfo{booktitle}{\emph{Proceedings of the 2019 Conference on Empirical
  Methods in Natural Language Processing and the 9th International Joint
  Conference on Natural Language Processing (EMNLP-IJCNLP)}}.
  \bibinfo{publisher}{Association for Computational Linguistics},
  \bibinfo{address}{Hong Kong, China}, \bibinfo{pages}{1664--1674}.
\newblock
\urldef\tempurl%
\url{https://doi.org/10.18653/v1/D19-1176}
\showDOI{\tempurl}


\bibitem[D'Alessio and Allen(2000)]%
        {dalessioMediaBiasPresidential2000a}
\bibfield{author}{\bibinfo{person}{Dave D'Alessio} {and} \bibinfo{person}{Mike
  Allen}.} \bibinfo{year}{2000}\natexlab{}.
\newblock \showarticletitle{Media {Bias} in {Presidential} {Elections}: {A}
  {Meta}-{Analysis}}.
\newblock \bibinfo{journal}{\emph{Journal of Communication}}
  \bibinfo{volume}{50}, \bibinfo{number}{4} (\bibinfo{date}{Dec.}
  \bibinfo{year}{2000}), \bibinfo{pages}{133--156}.
\newblock
\showISSN{0021-9916, 1460-2466}
\urldef\tempurl%
\url{https://doi.org/10.1111/j.1460-2466.2000.tb02866.x}
\showDOI{\tempurl}


\bibitem[Fair(1993)]%
        {fairWarFaminePoverty1993}
\bibfield{author}{\bibinfo{person}{Jo~Ellen Fair}.}
  \bibinfo{year}{1993}\natexlab{}.
\newblock \showarticletitle{War, {Famine}, and {Poverty}: {Race} in the
  {Construction} of {Africa}'s {Media} {Image}}.
\newblock \bibinfo{journal}{\emph{Journal of Communication Inquiry}}
  \bibinfo{volume}{17}, \bibinfo{number}{2} (\bibinfo{date}{July}
  \bibinfo{year}{1993}), \bibinfo{pages}{5--22}.
\newblock
\showISSN{0196-8599}
\urldef\tempurl%
\url{https://doi.org/10.1177/019685999301700202}
\showDOI{\tempurl}
\newblock
\shownote{Publisher: SAGE Publications Inc}.


\bibitem[Fan et~al\mbox{.}(2019)]%
        {fanPlainSightMedia2019}
\bibfield{author}{\bibinfo{person}{Lisa Fan}, \bibinfo{person}{Marshall White},
  \bibinfo{person}{Eva Sharma}, \bibinfo{person}{Ruisi Su},
  \bibinfo{person}{Prafulla~Kumar Choubey}, \bibinfo{person}{Ruihong Huang},
  {and} \bibinfo{person}{Lu Wang}.} \bibinfo{year}{2019}\natexlab{}.
\newblock \bibinfo{title}{In {Plain} {Sight}: {Media} {Bias} {Through} the
  {Lens} of {Factual} {Reporting}}.
\newblock
\newblock
\urldef\tempurl%
\url{https://doi.org/10.48550/arXiv.1909.02670}
\showDOI{\tempurl}
\newblock
\shownote{arXiv:1909.02670 [cs]}.


\bibitem[Feldman(2013)]%
        {feldmanPoliticalIdeology2013}
\bibfield{author}{\bibinfo{person}{Stanley Feldman}.}
  \bibinfo{year}{2013}\natexlab{}.
\newblock \showarticletitle{Political ideology}.
\newblock In \bibinfo{booktitle}{\emph{The {Oxford} handbook of political
  psychology, 2nd ed}}. \bibinfo{publisher}{Oxford University Press},
  \bibinfo{address}{New York, NY, US}, \bibinfo{pages}{591--626}.
\newblock
\showISBNx{978-0-19-976010-7}


\bibitem[Färber et~al\mbox{.}(2020)]%
        {farberMultidimensionalDatasetBased2020a}
\bibfield{author}{\bibinfo{person}{Michael Färber}, \bibinfo{person}{Victoria
  Burkard}, \bibinfo{person}{Adam Jatowt}, {and} \bibinfo{person}{Sora Lim}.}
  \bibinfo{year}{2020}\natexlab{}.
\newblock \showarticletitle{A {Multidimensional} {Dataset} {Based} on
  {Crowdsourcing} for {Analyzing} and {Detecting} {News} {Bias}}. In
  \bibinfo{booktitle}{\emph{Proceedings of the 29th {ACM} {International}
  {Conference} on {Information} \& {Knowledge} {Management}}}.
  \bibinfo{publisher}{ACM}, \bibinfo{address}{Virtual Event Ireland},
  \bibinfo{pages}{3007--3014}.
\newblock
\showISBNx{978-1-4503-6859-9}
\urldef\tempurl%
\url{https://doi.org/10.1145/3340531.3412876}
\showDOI{\tempurl}


\bibitem[Gerber et~al\mbox{.}(2009)]%
        {gerberDoesMediaMatter2009}
\bibfield{author}{\bibinfo{person}{Alan~S Gerber}, \bibinfo{person}{Dean
  Karlan}, {and} \bibinfo{person}{Daniel Bergan}.}
  \bibinfo{year}{2009}\natexlab{}.
\newblock \showarticletitle{Does the {Media} {Matter}? {A} {Field} {Experiment}
  {Measuring} the {Effect} of {Newspapers} on {Voting} {Behavior} and
  {Political} {Opinions}}.
\newblock \bibinfo{journal}{\emph{American Economic Journal: Applied
  Economics}} \bibinfo{volume}{1}, \bibinfo{number}{2} (\bibinfo{date}{March}
  \bibinfo{year}{2009}), \bibinfo{pages}{35--52}.
\newblock
\showISSN{1945-7782, 1945-7790}
\urldef\tempurl%
\url{https://doi.org/10.1257/app.1.2.35}
\showDOI{\tempurl}


\bibitem[Ghoshal(2018)]%
        {ghoshalRacialBiasTwitter2018}
\bibfield{author}{\bibinfo{person}{Torumoy Ghoshal}.}
  \bibinfo{year}{2018}\natexlab{}.
\newblock \bibinfo{title}{Racial {Bias} {Twitter}}.
\newblock
\newblock
\urldef\tempurl%
\url{https://github.com/tgh499/racial_bias_twitter}
\showURL{%
\tempurl}


\bibitem[Golbeck et~al\mbox{.}(2017)]%
        {golbeckLargeLabeledCorpus2017}
\bibfield{author}{\bibinfo{person}{Jennifer Golbeck}, \bibinfo{person}{Zahra
  Ashktorab}, \bibinfo{person}{Rashad~O. Banjo}, \bibinfo{person}{Alexandra
  Berlinger}, \bibinfo{person}{Siddharth Bhagwan}, \bibinfo{person}{Cody
  Buntain}, \bibinfo{person}{Paul Cheakalos}, \bibinfo{person}{Alicia~A.
  Geller}, \bibinfo{person}{Quint Gergory}, \bibinfo{person}{Rajesh~Kumar
  Gnanasekaran}, \bibinfo{person}{Raja~Rajan Gunasekaran},
  \bibinfo{person}{Kelly~M. Hoffman}, \bibinfo{person}{Jenny Hottle},
  \bibinfo{person}{Vichita Jienjitlert}, \bibinfo{person}{Shivika Khare},
  \bibinfo{person}{Ryan Lau}, \bibinfo{person}{Marianna~J. Martindale},
  \bibinfo{person}{Shalmali Naik}, \bibinfo{person}{Heather~L. Nixon},
  \bibinfo{person}{Piyush Ramachandran}, \bibinfo{person}{Kristine~M. Rogers},
  \bibinfo{person}{Lisa Rogers}, \bibinfo{person}{Meghna~Sardana Sarin},
  \bibinfo{person}{Gaurav Shahane}, \bibinfo{person}{Jayanee Thanki},
  \bibinfo{person}{Priyanka Vengataraman}, \bibinfo{person}{Zijian Wan}, {and}
  \bibinfo{person}{Derek~Michael Wu}.} \bibinfo{year}{2017}\natexlab{}.
\newblock \showarticletitle{A {Large} {Labeled} {Corpus} for {Online}
  {Harassment} {Research}}. In \bibinfo{booktitle}{\emph{Proceedings of the
  2017 {ACM} on {Web} {Science} {Conference}}}. \bibinfo{publisher}{ACM},
  \bibinfo{address}{Troy New York USA}, \bibinfo{pages}{229--233}.
\newblock
\showISBNx{978-1-4503-4896-6}
\urldef\tempurl%
\url{https://doi.org/10.1145/3091478.3091509}
\showDOI{\tempurl}


\bibitem[Grosz and Conde-Cespedes(2020)]%
        {groszAutomaticDetectionSexist2020}
\bibfield{author}{\bibinfo{person}{Dylan Grosz} {and} \bibinfo{person}{Patricia
  Conde-Cespedes}.} \bibinfo{year}{2020}\natexlab{}.
\newblock \showarticletitle{Automatic {Detection} of {Sexist} {Statements}
  {Commonly} {Used} at the {Workplace}}.
\newblock In \bibinfo{booktitle}{\emph{Trends and {Applications} in {Knowledge}
  {Discovery} and {Data} {Mining}}}, \bibfield{editor}{\bibinfo{person}{Wei Lu}
  {and} \bibinfo{person}{Kenny~Q. Zhu}} (Eds.). Vol.~\bibinfo{volume}{12237}.
  \bibinfo{publisher}{Springer International Publishing},
  \bibinfo{address}{Cham}, \bibinfo{pages}{104--115}.
\newblock
\showISBNx{978-3-030-60469-1 978-3-030-60470-7}
\urldef\tempurl%
\url{https://doi.org/10.1007/978-3-030-60470-7_11}
\showDOI{\tempurl}


\bibitem[Hofmann et~al\mbox{.}(2021)]%
        {Hofmann2021}
\bibfield{author}{\bibinfo{person}{Valentin Hofmann}, \bibinfo{person}{Xiaowen
  Dong}, \bibinfo{person}{Janet~B. Pierrehumbert}, {and}
  \bibinfo{person}{Hinrich Schütze}.} \bibinfo{year}{2021}\natexlab{}.
\newblock \bibinfo{title}{Modeling Ideological Salience and Framing in
  Polarized Online Groups with Graph Neural Networks and Structured Sparsity}.
\newblock
\newblock
\urldef\tempurl%
\url{https://doi.org/10.48550/ARXIV.2104.08829}
\showDOI{\tempurl}


\bibitem[Hube and Fetahu(2019a)]%
        {hubeNeuralBasedStatement2019}
\bibfield{author}{\bibinfo{person}{Christoph Hube} {and}
  \bibinfo{person}{Besnik Fetahu}.} \bibinfo{year}{2019}\natexlab{a}.
\newblock \showarticletitle{Neural {Based} {Statement} {Classification} for
  {Biased} {Language}}. In \bibinfo{booktitle}{\emph{Proceedings of the
  {Twelfth} {ACM} {International} {Conference} on {Web} {Search} and {Data}
  {Mining}}}. \bibinfo{publisher}{ACM}, \bibinfo{address}{Melbourne VIC
  Australia}, \bibinfo{pages}{195--203}.
\newblock
\showISBNx{978-1-4503-5940-5}
\urldef\tempurl%
\url{https://doi.org/10.1145/3289600.3291018}
\showDOI{\tempurl}


\bibitem[Hube and Fetahu(2019b)]%
        {hubeNeuralBasedStatement2019b}
\bibfield{author}{\bibinfo{person}{Christoph Hube} {and}
  \bibinfo{person}{Besnik Fetahu}.} \bibinfo{year}{2019}\natexlab{b}.
\newblock \showarticletitle{Neural {Based} {Statement} {Classification} for
  {Biased} {Language}}. In \bibinfo{booktitle}{\emph{Proceedings of the
  {Twelfth} {ACM} {International} {Conference} on {Web} {Search} and {Data}
  {Mining}}}. \bibinfo{publisher}{ACM}, \bibinfo{address}{Melbourne VIC
  Australia}, \bibinfo{pages}{195--203}.
\newblock
\showISBNx{978-1-4503-5940-5}
\urldef\tempurl%
\url{https://doi.org/10.1145/3289600.3291018}
\showDOI{\tempurl}


\bibitem[Huguet~Cabot et~al\mbox{.}(2021)]%
        {huguetcabotUsVsThem2021}
\bibfield{author}{\bibinfo{person}{Pere-Lluís Huguet~Cabot},
  \bibinfo{person}{David Abadi}, \bibinfo{person}{Agneta Fischer}, {and}
  \bibinfo{person}{Ekaterina Shutova}.} \bibinfo{year}{2021}\natexlab{}.
\newblock \showarticletitle{Us vs. {Them}: {A} {Dataset} of {Populist}
  {Attitudes}, {News} {Bias} and {Emotions}}. In
  \bibinfo{booktitle}{\emph{Proceedings of the 16th {Conference} of the
  {European} {Chapter} of the {Association} for {Computational} {Linguistics}:
  {Main} {Volume}}}. \bibinfo{publisher}{Association for Computational
  Linguistics}, \bibinfo{address}{Online}, \bibinfo{pages}{1921--1945}.
\newblock
\urldef\tempurl%
\url{https://doi.org/10.18653/v1/2021.eacl-main.165}
\showDOI{\tempurl}


\bibitem[Kiesel et~al\mbox{.}(2018)]%
        {kieselDataPANSemEval2018}
\bibfield{author}{\bibinfo{person}{Johannes Kiesel}, \bibinfo{person}{Maria
  Mestre}, \bibinfo{person}{Rishabh Shukla}, \bibinfo{person}{Emmanuel
  Vincent}, \bibinfo{person}{David Corney}, \bibinfo{person}{Payam Adineh},
  \bibinfo{person}{Benno Stein}, {and} \bibinfo{person}{Martin Potthast}.}
  \bibinfo{year}{2018}\natexlab{}.
\newblock \bibinfo{title}{Data for {PAN} at {SemEval} 2019 {Task} 4:
  {Hyperpartisan} {News} {Detection}}.
\newblock
\newblock
\urldef\tempurl%
\url{https://doi.org/10.5281/ZENODO.1489920}
\showDOI{\tempurl}
\newblock
\shownote{Version Number: Training and validation v1 Type: dataset}.


\bibitem[Krieger et~al\mbox{.}(2022)]%
        {krieger_domain-adaptive_2022}
\bibfield{author}{\bibinfo{person}{David Krieger}, \bibinfo{person}{Timo
  Spinde}, \bibinfo{person}{Terry Ruas}, \bibinfo{person}{Juhi Kulshrestha},
  {and} \bibinfo{person}{Bela Gipp}.} \bibinfo{year}{2022}\natexlab{}.
\newblock \showarticletitle{A Domain-adaptive Pre-training Approach for
  Language Bias Detection in News}. In \bibinfo{booktitle}{\emph{2022 ACM/IEEE
  Joint Conference on Digital Libraries (JCDL)}} (2022-06-01).
  \bibinfo{address}{Cologne, Germany}.
\newblock
\urldef\tempurl%
\url{https://doi.org/10.1145/3529372.3530932}
\showDOI{\tempurl}


\bibitem[Kwak et~al\mbox{.}(2020)]%
        {kwakSystematicMediaFrame2020}
\bibfield{author}{\bibinfo{person}{Haewoon Kwak}, \bibinfo{person}{Jisun An},
  {and} \bibinfo{person}{Yong-Yeol Ahn}.} \bibinfo{year}{2020}\natexlab{}.
\newblock \showarticletitle{A {Systematic} {Media} {Frame} {Analysis} of 1.5
  {Million} {New} {York} {Times} {Articles} from 2000 to 2017}. In
  \bibinfo{booktitle}{\emph{12th {ACM} {Conference} on {Web} {Science}}}.
  \bibinfo{publisher}{ACM}, \bibinfo{address}{Southampton United Kingdom},
  \bibinfo{pages}{305--314}.
\newblock
\showISBNx{978-1-4503-7989-2}
\urldef\tempurl%
\url{https://doi.org/10.1145/3394231.3397921}
\showDOI{\tempurl}


\bibitem[Lavery(2013)]%
        {laveryGenderBiasMedia2013}
\bibfield{author}{\bibinfo{person}{Lesley Lavery}.}
  \bibinfo{year}{2013}\natexlab{}.
\newblock \showarticletitle{Gender {Bias} in the {Media}? {An} {Examination} of
  {Local} {Television} {News} {Coverage} of {Male} and {Female} {House}
  {Candidates}: {Gender} {Bias} in the {Media}}.
\newblock \bibinfo{journal}{\emph{Politics \& Policy}} \bibinfo{volume}{41},
  \bibinfo{number}{6} (\bibinfo{date}{Dec.} \bibinfo{year}{2013}),
  \bibinfo{pages}{877--910}.
\newblock
\showISSN{15555623}
\urldef\tempurl%
\url{https://doi.org/10.1111/polp.12051}
\showDOI{\tempurl}


\bibitem[Lee et~al\mbox{.}(2021)]%
        {Lee_2021}
\bibfield{author}{\bibinfo{person}{Nayeon Lee}, \bibinfo{person}{Yejin Bang},
  \bibinfo{person}{Andrea Madotto}, {and} \bibinfo{person}{Pascale Fung}.}
  \bibinfo{year}{2021}\natexlab{}.
\newblock \showarticletitle{Mitigating Media Bias through Neutral Article
  Generation}.
\newblock \bibinfo{journal}{\emph{CoRR}}  \bibinfo{volume}{abs/2104.00336}
  (\bibinfo{year}{2021}).
\newblock
\urldef\tempurl%
\url{https://doi.org/10.48550/arXiv.2104.00336}
\showDOI{\tempurl}


\bibitem[Lim et~al\mbox{.}(2020)]%
        {zotero-1084}
\bibfield{author}{\bibinfo{person}{Sora Lim}, \bibinfo{person}{Adam Jatowt},
  {and} \bibinfo{person}{Y Masatoshi}.} \bibinfo{year}{2020}\natexlab{}.
\newblock \showarticletitle{Creating a dataset for fine-grained bias detection
  in news articles} \emph{(\bibinfo{series}{12})}. \bibinfo{pages}{1--35}.
\newblock


\bibitem[Liu et~al\mbox{.}(2019)]%
        {liuRoBERTaRobustlyOptimized2019}
\bibfield{author}{\bibinfo{person}{Yinhan Liu}, \bibinfo{person}{Myle Ott},
  \bibinfo{person}{Naman Goyal}, \bibinfo{person}{Jingfei Du},
  \bibinfo{person}{Mandar Joshi}, \bibinfo{person}{Danqi Chen},
  \bibinfo{person}{Omer Levy}, \bibinfo{person}{Mike Lewis},
  \bibinfo{person}{Luke Zettlemoyer}, {and} \bibinfo{person}{Veselin
  Stoyanov}.} \bibinfo{year}{2019}\natexlab{}.
\newblock \showarticletitle{{RoBERTa}: {A} {Robustly} {Optimized} {BERT}
  {Pretraining} {Approach}}.
\newblock  (\bibinfo{year}{2019}).
\newblock
\urldef\tempurl%
\url{https://doi.org/10.48550/ARXIV.1907.11692}
\showDOI{\tempurl}
\newblock
\shownote{Publisher: arXiv Version Number: 1}.


\bibitem[Liu et~al\mbox{.}(2022)]%
        {liuPOLITICSPretrainingSamestory2022}
\bibfield{author}{\bibinfo{person}{Yujian Liu},
  \bibinfo{person}{Xinliang~Frederick Zhang}, \bibinfo{person}{David Wegsman},
  \bibinfo{person}{Nick Beauchamp}, {and} \bibinfo{person}{Lu Wang}.}
  \bibinfo{year}{2022}\natexlab{}.
\newblock \showarticletitle{{POLITICS}: {Pretraining} with {Same}-story
  {Article} {Comparison} for {Ideology} {Prediction} and {Stance} {Detection}}.
\newblock  (\bibinfo{year}{2022}).
\newblock
\urldef\tempurl%
\url{https://doi.org/10.48550/ARXIV.2205.00619}
\showDOI{\tempurl}
\newblock
\shownote{Publisher: arXiv Version Number: 1}.


\bibitem[Mathew et~al\mbox{.}(2021)]%
        {mathewHateXplainBenchmarkDataset2021a}
\bibfield{author}{\bibinfo{person}{Binny Mathew}, \bibinfo{person}{Punyajoy
  Saha}, \bibinfo{person}{Seid~Muhie Yimam}, \bibinfo{person}{Chris Biemann},
  \bibinfo{person}{Pawan Goyal}, {and} \bibinfo{person}{Animesh Mukherjee}.}
  \bibinfo{year}{2021}\natexlab{}.
\newblock \showarticletitle{{HateXplain}: {A} {Benchmark} {Dataset} for
  {Explainable} {Hate} {Speech} {Detection}}.
\newblock \bibinfo{journal}{\emph{Proceedings of the AAAI Conference on
  Artificial Intelligence}} \bibinfo{volume}{35}, \bibinfo{number}{17}
  (\bibinfo{date}{May} \bibinfo{year}{2021}), \bibinfo{pages}{14867--14875}.
\newblock
\showISSN{2374-3468, 2159-5399}
\urldef\tempurl%
\url{https://doi.org/10.1609/aaai.v35i17.17745}
\showDOI{\tempurl}


\bibitem[Min and Feaster(2010)]%
        {minMissingChildrenNational2010}
\bibfield{author}{\bibinfo{person}{Seong-Jae Min} {and}
  \bibinfo{person}{John~C. Feaster}.} \bibinfo{year}{2010}\natexlab{}.
\newblock \showarticletitle{Missing {Children} in {National} {News} {Coverage}:
  {Racial} and {Gender} {Representations} of {Missing} {Children} {Cases}}.
\newblock \bibinfo{journal}{\emph{Communication Research Reports}}
  \bibinfo{volume}{27}, \bibinfo{number}{3} (\bibinfo{date}{Aug.}
  \bibinfo{year}{2010}), \bibinfo{pages}{207--216}.
\newblock
\showISSN{0882-4096}
\urldef\tempurl%
\url{https://doi.org/10.1080/08824091003776289}
\showDOI{\tempurl}
\newblock
\shownote{Publisher: Routledge \_eprint:
  https://doi.org/10.1080/08824091003776289}.


\bibitem[Mozafari et~al\mbox{.}(2020)]%
        {mozafari2020hate}
\bibfield{author}{\bibinfo{person}{Marzieh Mozafari}, \bibinfo{person}{Reza
  Farahbakhsh}, {and} \bibinfo{person}{No{\"e}l Crespi}.}
  \bibinfo{year}{2020}\natexlab{}.
\newblock \showarticletitle{Hate speech detection and racial bias mitigation in
  social media based on BERT model}.
\newblock \bibinfo{journal}{\emph{PloS one}} \bibinfo{volume}{15},
  \bibinfo{number}{8} (\bibinfo{year}{2020}), \bibinfo{pages}{e0237861}.
\newblock
\urldef\tempurl%
\url{https://doi.org/10.1371/journal.pone.0237861}
\showDOI{\tempurl}


\bibitem[Nickerson(1998)]%
        {nickersonConfirmationBiasUbiquitous1998}
\bibfield{author}{\bibinfo{person}{Raymond~S. Nickerson}.}
  \bibinfo{year}{1998}\natexlab{}.
\newblock \showarticletitle{Confirmation {Bias}: {A} {Ubiquitous} {Phenomenon}
  in {Many} {Guises}}.
\newblock \bibinfo{journal}{\emph{Review of General Psychology}}
  \bibinfo{volume}{2}, \bibinfo{number}{2} (\bibinfo{date}{June}
  \bibinfo{year}{1998}), \bibinfo{pages}{175--220}.
\newblock
\showISSN{1089-2680, 1939-1552}
\urldef\tempurl%
\url{https://doi.org/10.1037/1089-2680.2.2.175}
\showDOI{\tempurl}


\bibitem[Potts et~al\mbox{.}(2021)]%
        {potts-etal-2021-dynasent}
\bibfield{author}{\bibinfo{person}{Christopher Potts},
  \bibinfo{person}{Zhengxuan Wu}, \bibinfo{person}{Atticus Geiger}, {and}
  \bibinfo{person}{Douwe Kiela}.} \bibinfo{year}{2021}\natexlab{}.
\newblock \showarticletitle{{D}yna{S}ent: A Dynamic Benchmark for Sentiment
  Analysis}. In \bibinfo{booktitle}{\emph{Proceedings of the 59th Annual
  Meeting of the Association for Computational Linguistics and the 11th
  International Joint Conference on Natural Language Processing (Volume 1: Long
  Papers)}}. \bibinfo{publisher}{Association for Computational Linguistics},
  \bibinfo{address}{Online}, \bibinfo{pages}{2388--2404}.
\newblock
\urldef\tempurl%
\url{https://doi.org/10.18653/v1/2021.acl-long.186}
\showDOI{\tempurl}


\bibitem[Pryzant et~al\mbox{.}(2019)]%
        {pryzantAutomaticallyNeutralizingSubjective2019}
\bibfield{author}{\bibinfo{person}{Reid Pryzant},
  \bibinfo{person}{Richard~Diehl Martinez}, \bibinfo{person}{Nathan Dass},
  \bibinfo{person}{Sadao Kurohashi}, \bibinfo{person}{Dan Jurafsky}, {and}
  \bibinfo{person}{Diyi Yang}.} \bibinfo{year}{2019}\natexlab{}.
\newblock \showarticletitle{Automatically {Neutralizing} {Subjective} {Bias} in
  {Text}}.
\newblock  (\bibinfo{year}{2019}).
\newblock
\urldef\tempurl%
\url{https://doi.org/10.48550/ARXIV.1911.09709}
\showDOI{\tempurl}
\newblock
\shownote{Publisher: arXiv Version Number: 3}.


\bibitem[Recasens et~al\mbox{.}(2013)]%
        {recasensLinguisticModelsAnalyzing2013}
\bibfield{author}{\bibinfo{person}{Marta Recasens}, \bibinfo{person}{Cristian
  Danescu-Niculescu-Mizil}, {and} \bibinfo{person}{Dan Jurafsky}.}
  \bibinfo{year}{2013}\natexlab{}.
\newblock \showarticletitle{Linguistic {Models} for {Analyzing} and {Detecting}
  {Biased} {Language}}. In \bibinfo{booktitle}{\emph{Proceedings of the 51st
  {Annual} {Meeting} of the {Association} for {Computational} {Linguistics}
  ({Volume} 1: {Long} {Papers})}}. \bibinfo{publisher}{Association for
  Computational Linguistics}, \bibinfo{address}{Sofia, Bulgaria},
  \bibinfo{pages}{1650--1659}.
\newblock
\urldef\tempurl%
\url{https://aclanthology.org/P13-1162}
\showURL{%
\tempurl}


\bibitem[Ribeiro et~al\mbox{.}(2020)]%
        {ribeiro-etal-2020-beyond}
\bibfield{author}{\bibinfo{person}{Marco~Tulio Ribeiro},
  \bibinfo{person}{Tongshuang Wu}, \bibinfo{person}{Carlos Guestrin}, {and}
  \bibinfo{person}{Sameer Singh}.} \bibinfo{year}{2020}\natexlab{}.
\newblock \showarticletitle{Beyond Accuracy: Behavioral Testing of {NLP} Models
  with {C}heck{L}ist}. In \bibinfo{booktitle}{\emph{Proceedings of the 58th
  Annual Meeting of the Association for Computational Linguistics}}.
  \bibinfo{publisher}{Association for Computational Linguistics},
  \bibinfo{address}{Online}, \bibinfo{pages}{4902--4912}.
\newblock
\urldef\tempurl%
\url{https://doi.org/10.18653/v1/2020.acl-main.442}
\showDOI{\tempurl}


\bibitem[Rocha et~al\mbox{.}(2021)]%
        {rochaImpactFakeNews2021}
\bibfield{author}{\bibinfo{person}{Yasmim~Mendes Rocha},
  \bibinfo{person}{Gabriel~Acácio de Moura}, \bibinfo{person}{Gabriel~Alves
  Desidério}, \bibinfo{person}{Carlos~Henrique de Oliveira},
  \bibinfo{person}{Francisco~Dantas Lourenço}, {and}
  \bibinfo{person}{Larissa~Deadame de Figueiredo~Nicolete}.}
  \bibinfo{year}{2021}\natexlab{}.
\newblock \showarticletitle{The impact of fake news on social media and its
  influence on health during the {COVID}-19 pandemic: a systematic review}.
\newblock \bibinfo{journal}{\emph{Journal of Public Health}}
  (\bibinfo{date}{Oct.} \bibinfo{year}{2021}).
\newblock
\showISSN{1613-2238}
\urldef\tempurl%
\url{https://doi.org/10.1007/s10389-021-01658-z}
\showDOI{\tempurl}


\bibitem[Samory et~al\mbox{.}(2020)]%
        {samoryCallMeSexist2020}
\bibfield{author}{\bibinfo{person}{Mattia Samory}, \bibinfo{person}{Indira
  Sen}, \bibinfo{person}{Julian Kohne}, \bibinfo{person}{Fabian Floeck}, {and}
  \bibinfo{person}{Claudia Wagner}.} \bibinfo{year}{2020}\natexlab{}.
\newblock \showarticletitle{"{Call} me sexist, but...": {Revisiting} {Sexism}
  {Detection} {Using} {Psychological} {Scales} and {Adversarial} {Samples}}.
\newblock  (\bibinfo{year}{2020}).
\newblock
\urldef\tempurl%
\url{https://doi.org/10.48550/ARXIV.2004.12764}
\showDOI{\tempurl}


\bibitem[Shu et~al\mbox{.}(2020)]%
        {shuFakeNewsNetDataRepository2020}
\bibfield{author}{\bibinfo{person}{Kai Shu}, \bibinfo{person}{Deepak
  Mahudeswaran}, \bibinfo{person}{Suhang Wang}, \bibinfo{person}{Dongwon Lee},
  {and} \bibinfo{person}{Huan Liu}.} \bibinfo{year}{2020}\natexlab{}.
\newblock \showarticletitle{{FakeNewsNet}: {A} {Data} {Repository} with {News}
  {Content}, {Social} {Context}, and {Spatiotemporal} {Information} for
  {Studying} {Fake} {News} on {Social} {Media}}.
\newblock \bibinfo{journal}{\emph{Big Data}} \bibinfo{volume}{8},
  \bibinfo{number}{3} (\bibinfo{date}{June} \bibinfo{year}{2020}),
  \bibinfo{pages}{171--188}.
\newblock
\showISSN{2167-6461, 2167-647X}
\urldef\tempurl%
\url{https://doi.org/10.1089/big.2020.0062}
\showDOI{\tempurl}


\bibitem[Singh et~al\mbox{.}(2020)]%
        {singhFemaleLibrariansMale2020a}
\bibfield{author}{\bibinfo{person}{Vivek~K. Singh}, \bibinfo{person}{Mary
  Chayko}, \bibinfo{person}{Raj Inamdar}, {and} \bibinfo{person}{Diana
  Floegel}.} \bibinfo{year}{2020}\natexlab{}.
\newblock \showarticletitle{Female librarians and male computer programmers?
  {Gender} bias in occupational images on digital media platforms}.
\newblock \bibinfo{journal}{\emph{Journal of the Association for Information
  Science and Technology}} \bibinfo{volume}{71}, \bibinfo{number}{11}
  (\bibinfo{date}{Nov.} \bibinfo{year}{2020}), \bibinfo{pages}{1281--1294}.
\newblock
\showISSN{2330-1635, 2330-1643}
\urldef\tempurl%
\url{https://doi.org/10.1002/asi.24335}
\showDOI{\tempurl}


\bibitem[Spinde({[n.\,d.]})]%
        {spinde-interdisciplinary-2021}
\bibfield{author}{\bibinfo{person}{Timo Spinde}.}
  \bibinfo{year}{[n.\,d.]}\natexlab{}.
\newblock \showarticletitle{An Interdisciplinary Approach for the Automated
  Detection and Visualization of Media Bias in News Articles}. In
  \bibinfo{booktitle}{\emph{2021 {IEEE} International Conference on Data Mining
  Workshops ({ICDMW})}} (2021-09-30).
\newblock
\urldef\tempurl%
\url{https://doi.org/10.1109/ICDMW53433.2021.00144}
\showDOI{\tempurl}


\bibitem[Spinde et~al\mbox{.}(2020a)]%
        {spinde-integrated-2020}
\bibfield{author}{\bibinfo{person}{Timo Spinde}, \bibinfo{person}{Felix
  Hamborg}, {and} \bibinfo{person}{Bela Gipp}.}
  \bibinfo{year}{2020}\natexlab{a}.
\newblock \showarticletitle{An Integrated Approach to Detect Media Bias in
  German News Articles}. In \bibinfo{booktitle}{\emph{Proceedings of the
  ACM/IEEE Joint Conference on Digital Libraries in 2020}} (2020-01-01)
  \emph{(\bibinfo{series}{JCDL '20})}. \bibinfo{publisher}{Association for
  Computing Machinery}, \bibinfo{address}{Virtual Event, China},
  \bibinfo{pages}{505–506}.
\newblock
\showISBNx{9781450375856}
\urldef\tempurl%
\url{https://doi.org/10.1145/3383583.3398585}
\showDOI{\tempurl}


\bibitem[Spinde et~al\mbox{.}(2020b)]%
        {SpindeGerman2020}
\bibfield{author}{\bibinfo{person}{Timo Spinde}, \bibinfo{person}{Felix
  Hamborg}, {and} \bibinfo{person}{Bela Gipp}.}
  \bibinfo{year}{2020}\natexlab{b}.
\newblock \showarticletitle{Media Bias in German News Articles: A Combined
  Approach}. In \bibinfo{booktitle}{\emph{ECML PKDD 2020 Workshops}},
  \bibfield{editor}{\bibinfo{person}{Irena Koprinska}, \bibinfo{person}{Michael
  Kamp}, \bibinfo{person}{Annalisa Appice}, \bibinfo{person}{Corrado Loglisci},
  \bibinfo{person}{Luiza Antonie}, \bibinfo{person}{Albrecht Zimmermann},
  \bibinfo{person}{Riccardo Guidotti}, \bibinfo{person}{{\"O}zlem
  {\"O}zg{\"o}bek}, \bibinfo{person}{Rita~P. Ribeiro}, \bibinfo{person}{Ricard
  Gavald{\`a}}, \bibinfo{person}{Jo{\~a}o Gama}, \bibinfo{person}{Linara
  Adilova}, \bibinfo{person}{Yamuna Krishnamurthy}, \bibinfo{person}{Pedro~M.
  Ferreira}, \bibinfo{person}{Donato Malerba}, \bibinfo{person}{Ib{\'e}ria
  Medeiros}, \bibinfo{person}{Michelangelo Ceci}, \bibinfo{person}{Giuseppe
  Manco}, \bibinfo{person}{Elio Masciari}, \bibinfo{person}{Zbigniew~W. Ras},
  \bibinfo{person}{Peter Christen}, \bibinfo{person}{Eirini Ntoutsi},
  \bibinfo{person}{Erich Schubert}, \bibinfo{person}{Arthur Zimek},
  \bibinfo{person}{Anna Monreale}, \bibinfo{person}{Przemyslaw Biecek},
  \bibinfo{person}{Salvatore Rinzivillo}, \bibinfo{person}{Benjamin Kille},
  \bibinfo{person}{Andreas Lommatzsch}, {and} \bibinfo{person}{Jon~Atle Gulla}}
  (Eds.). \bibinfo{publisher}{Springer International Publishing},
  \bibinfo{address}{Cham}, \bibinfo{pages}{581--590}.
\newblock
\showISBNx{978-3-030-65965-3}


\bibitem[Spinde et~al\mbox{.}(2023)]%
        {spindeIntroducingMediaBias2022}
\bibfield{author}{\bibinfo{person}{Timo Spinde}, \bibinfo{person}{Smi
  Hinterreiter}, \bibinfo{person}{Fabian Haak}, \bibinfo{person}{Terry Ruas},
  \bibinfo{person}{Helge Giese}, \bibinfo{person}{Norman Meuschke}, {and}
  \bibinfo{person}{Bela Gipp}.} \bibinfo{year}{2023}\natexlab{}.
\newblock \showarticletitle{The Media Bias Taxonomy: A Systematic Literature
  Review on the Forms and Automated Detection of Media Bias}.
\newblock \bibinfo{journal}{\emph{CSUR}} (\bibinfo{year}{2023}).
\newblock
\newblock
\shownote{[in review]}.


\bibitem[Spinde et~al\mbox{.}(2022a)]%
        {spinde-how-2021}
\bibfield{author}{\bibinfo{person}{Timo Spinde}, \bibinfo{person}{Christin
  Jeggle}, \bibinfo{person}{Magdalena Haupt}, \bibinfo{person}{Wolfgang
  Gaissmaier}, {and} \bibinfo{person}{Helge Giese}.}
  \bibinfo{year}{2022}\natexlab{a}.
\newblock \showarticletitle{How do we raise media bias awareness effectively?
  Effects of visualizations to communicate bias}.
\newblock \bibinfo{journal}{\emph{PLOS ONE}} \bibinfo{volume}{17},
  \bibinfo{number}{4}, \bibinfo{pages}{1--14}.
\newblock
\urldef\tempurl%
\url{https://doi.org/10.1371/journal.pone.0266204}
\showDOI{\tempurl}


\bibitem[Spinde et~al\mbox{.}(2021a)]%
        {spinde-you-2021}
\bibfield{author}{\bibinfo{person}{Timo Spinde}, \bibinfo{person}{Christina
  Kreuter}, \bibinfo{person}{Wolfgang Gaissmaier}, \bibinfo{person}{Felix
  Hamborg}, \bibinfo{person}{Bela Gipp}, {and} \bibinfo{person}{Helge Giese}.}
  \bibinfo{year}{2021}\natexlab{a}.
\newblock \showarticletitle{Do You Think It’s Biased? How To Ask For The
  Perception Of Media Bias}. In \bibinfo{booktitle}{\emph{Proceedings of the
  ACM/IEEE Joint Conference on Digital Libraries (JCDL)}} (2021-09-01).
\newblock
\urldef\tempurl%
\url{https://doi.org/10.1109/JCDL52503.2021.00018}
\showDOI{\tempurl}


\bibitem[Spinde et~al\mbox{.}(2021b)]%
        {spinde-towards-2021}
\bibfield{author}{\bibinfo{person}{Timo Spinde}, \bibinfo{person}{David
  Krieger}, \bibinfo{person}{Manu Plank}, {and} \bibinfo{person}{Bela Gipp}.}
  \bibinfo{year}{2021}\natexlab{b}.
\newblock \showarticletitle{Towards A Reliable Ground-Truth For Biased Language
  Detection}. In \bibinfo{booktitle}{\emph{Proceedings of the ACM/IEEE-CS Joint
  Conference on Digital Libraries (JCDL)}} (2021-09-01).
  \bibinfo{address}{Virtual Event}.
\newblock
\urldef\tempurl%
\url{https://doi.org/10.1109/JCDL52503.2021.00053}
\showDOI{\tempurl}


\bibitem[Spinde et~al\mbox{.}(2022b)]%
        {spinde-exploiting-2021}
\bibfield{author}{\bibinfo{person}{Timo Spinde}, \bibinfo{person}{Jan-David
  Krieger}, \bibinfo{person}{Terry Ruas}, \bibinfo{person}{Jelena Mitrović},
  \bibinfo{person}{Franz Götz-Hahn}, \bibinfo{person}{Akiko Aizawa}, {and}
  \bibinfo{person}{Bela Gipp}.} \bibinfo{year}{2022}\natexlab{b}.
\newblock \showarticletitle{Exploiting Transformer-based Multitask Learning for
  the Detection of Media Bias in News Articles}. In
  \bibinfo{booktitle}{\emph{Proceedings of the iConference 2022}} (2022-03-04).
  \bibinfo{address}{Virtual event}.
\newblock
\urldef\tempurl%
\url{https://doi.org/10.1007/978-3-030-96957-8_20}
\showDOI{\tempurl}


\bibitem[Spinde et~al\mbox{.}(2021c)]%
        {spindeNeuralMediaBias2021}
\bibfield{author}{\bibinfo{person}{Timo Spinde}, \bibinfo{person}{Manuel
  Plank}, \bibinfo{person}{Jan-David Krieger}, \bibinfo{person}{Terry Ruas},
  \bibinfo{person}{Bela Gipp}, {and} \bibinfo{person}{Akiko Aizawa}.}
  \bibinfo{year}{2021}\natexlab{c}.
\newblock \showarticletitle{Neural {Media} {Bias} {Detection} {Using} {Distant}
  {Supervision} {With} {BABE} - {Bias} {Annotations} {By} {Experts}}. In
  \bibinfo{booktitle}{\emph{Findings of the {Association} for {Computational}
  {Linguistics}: {EMNLP} 2021}}. \bibinfo{address}{Dominican Republic}.
\newblock
\urldef\tempurl%
\url{https://doi.org/10.18653/v1/2021.findings-emnlp.101}
\showDOI{\tempurl}


\bibitem[Spinde et~al\mbox{.}({[n.\,d.]})]%
        {spinde-tassy-2021}
\bibfield{author}{\bibinfo{person}{Timo Spinde}, \bibinfo{person}{Kanishka
  Sinha}, \bibinfo{person}{Norman Meuschke}, {and} \bibinfo{person}{Bela
  Gipp}.} \bibinfo{year}{[n.\,d.]}\natexlab{}.
\newblock \showarticletitle{{TASSY} - A Text Annotation Survey System}. In
  \bibinfo{booktitle}{\emph{Proceedings of the {ACM}/{IEEE} Joint Conference on
  Digital Libraries ({JCDL})}} (2021-09-01).
\newblock
\urldef\tempurl%
\url{https://doi.org/10.1109/JCDL52503.2021.00052}
\showDOI{\tempurl}


\bibitem[Srivastava et~al\mbox{.}(2022)]%
        {srivastavaImitationGameQuantifying2022}
\bibfield{author}{\bibinfo{person}{Aarohi Srivastava}, \bibinfo{person}{Abhinav
  Rastogi}, \bibinfo{person}{Abhishek Rao}, {and} \bibinfo{person}{et~al.
  Shoeb}.} \bibinfo{year}{2022}\natexlab{}.
\newblock \showarticletitle{Beyond the {Imitation} {Game}: {Quantifying} and
  extrapolating the capabilities of language models}.
\newblock  (\bibinfo{year}{2022}).
\newblock
\urldef\tempurl%
\url{https://doi.org/10.48550/ARXIV.2206.04615}
\showDOI{\tempurl}
\newblock
\shownote{Publisher: arXiv Version Number: 2}.


\bibitem[Tandoc~Jr.(2019)]%
        {tandocjr.FactsFakeNews2019}
\bibfield{author}{\bibinfo{person}{Edson~C. Tandoc~Jr.}}
  \bibinfo{year}{2019}\natexlab{}.
\newblock \showarticletitle{The facts of fake news: {A} research review}.
\newblock \bibinfo{journal}{\emph{Sociology Compass}} \bibinfo{volume}{13},
  \bibinfo{number}{9} (\bibinfo{year}{2019}), \bibinfo{pages}{e12724}.
\newblock
\showISSN{1751-9020}
\urldef\tempurl%
\url{https://doi.org/10.1111/soc4.12724}
\showDOI{\tempurl}
\newblock
\shownote{\_eprint:
  https://onlinelibrary.wiley.com/doi/pdf/10.1111/soc4.12724}.


\bibitem[Van~der Pas and Aaldering(2020)]%
        {vanderpas2020}
\bibfield{author}{\bibinfo{person}{Daphne~Joanna Van~der Pas} {and}
  \bibinfo{person}{Loes Aaldering}.} \bibinfo{year}{2020}\natexlab{}.
\newblock \showarticletitle{{Gender Differences in Political Media Coverage: A
  Meta-Analysis}}.
\newblock \bibinfo{journal}{\emph{Journal of Communication}}
  \bibinfo{volume}{70}, \bibinfo{number}{1} (\bibinfo{date}{02}
  \bibinfo{year}{2020}), \bibinfo{pages}{114--143}.
\newblock
\showISSN{0021-9916}
\urldef\tempurl%
\url{https://doi.org/10.1093/joc/jqz046}
\showDOI{\tempurl}
\showeprint{https://academic.oup.com/joc/article-pdf/70/1/114/34053729/jqz046.pdf}


\bibitem[Vidgen et~al\mbox{.}(2021)]%
        {vidgenIntroducingCADContextual2021}
\bibfield{author}{\bibinfo{person}{Bertie Vidgen}, \bibinfo{person}{Dong
  Nguyen}, \bibinfo{person}{Helen Margetts}, \bibinfo{person}{Patricia
  Rossini}, {and} \bibinfo{person}{Rebekah Tromble}.}
  \bibinfo{year}{2021}\natexlab{}.
\newblock \showarticletitle{Introducing {CAD}: the {Contextual} {Abuse}
  {Dataset}}. In \bibinfo{booktitle}{\emph{Proceedings of the 2021 {Conference}
  of the {North} {American} {Chapter} of the {Association} for {Computational}
  {Linguistics}: {Human} {Language} {Technologies}}}.
  \bibinfo{publisher}{Association for Computational Linguistics},
  \bibinfo{address}{Online}, \bibinfo{pages}{2289--2303}.
\newblock
\urldef\tempurl%
\url{https://doi.org/10.18653/v1/2021.naacl-main.182}
\showDOI{\tempurl}


\bibitem[Voigt et~al\mbox{.}(2018)]%
        {voigtRtGenderCorpusStudying2018}
\bibfield{author}{\bibinfo{person}{Rob Voigt}, \bibinfo{person}{David Jurgens},
  \bibinfo{person}{Vinodkumar Prabhakaran}, \bibinfo{person}{Dan Jurafsky},
  {and} \bibinfo{person}{Yulia Tsvetkov}.} \bibinfo{year}{2018}\natexlab{}.
\newblock \showarticletitle{{RtGender}: {A} {Corpus} for {Studying}
  {Differential} {Responses} to {Gender}}. In
  \bibinfo{booktitle}{\emph{Proceedings of the {Eleventh} {International}
  {Conference} on {Language} {Resources} and {Evaluation} ({LREC} 2018)}}.
  \bibinfo{publisher}{European Language Resources Association (ELRA)},
  \bibinfo{address}{Miyazaki, Japan}.
\newblock
\urldef\tempurl%
\url{https://aclanthology.org/L18-1445}
\showURL{%
\tempurl}


\bibitem[Wang et~al\mbox{.}(2019a)]%
        {wangSuperGLUEStickierBenchmark2019}
\bibfield{author}{\bibinfo{person}{Alex Wang}, \bibinfo{person}{Yada
  Pruksachatkun}, \bibinfo{person}{Nikita Nangia}, \bibinfo{person}{Amanpreet
  Singh}, \bibinfo{person}{Julian Michael}, \bibinfo{person}{Felix Hill},
  \bibinfo{person}{Omer Levy}, {and} \bibinfo{person}{Samuel Bowman}.}
  \bibinfo{year}{2019}\natexlab{a}.
\newblock \showarticletitle{{SuperGLUE}: {A} {Stickier} {Benchmark} for
  {General}-{Purpose} {Language} {Understanding} {Systems}}. In
  \bibinfo{booktitle}{\emph{Advances in {Neural} {Information} {Processing}
  {Systems}}}, Vol.~\bibinfo{volume}{32}. \bibinfo{publisher}{Curran
  Associates, Inc.}
\newblock
\urldef\tempurl%
\url{https://proceedings.neurips.cc/paper/2019/hash/4496bf24afe7fab6f046bf4923da8de6-Abstract.html}
\showURL{%
\tempurl}


\bibitem[Wang et~al\mbox{.}(2019b)]%
        {wangGLUEMultiTaskBenchmark2019}
\bibfield{author}{\bibinfo{person}{Alex Wang}, \bibinfo{person}{Amanpreet
  Singh}, \bibinfo{person}{Julian Michael}, \bibinfo{person}{Felix Hill},
  \bibinfo{person}{Omer Levy}, {and} \bibinfo{person}{Samuel~R. Bowman}.}
  \bibinfo{year}{2019}\natexlab{b}.
\newblock \bibinfo{title}{{GLUE}: {A} {Multi}-{Task} {Benchmark} and {Analysis}
  {Platform} for {Natural} {Language} {Understanding}}.
\newblock
\newblock
\urldef\tempurl%
\url{http://arxiv.org/abs/1804.07461}
\showURL{%
\tempurl}
\newblock
\shownote{arXiv:1804.07461 [cs]}.


\bibitem[Wang(2017)]%
        {wangLiarLiarPants2017a}
\bibfield{author}{\bibinfo{person}{William~Yang Wang}.}
  \bibinfo{year}{2017}\natexlab{}.
\newblock \showarticletitle{"{Liar}, {Liar} {Pants} on {Fire}": {A} {New}
  {Benchmark} {Dataset} for {Fake} {News} {Detection}}. In
  \bibinfo{booktitle}{\emph{Proceedings of the 55th {Annual} {Meeting} of the
  {Association} for {Computational} {Linguistics} ({Volume} 2: {Short}
  {Papers})}}. \bibinfo{publisher}{Association for Computational Linguistics},
  \bibinfo{address}{Vancouver, Canada}, \bibinfo{pages}{422--426}.
\newblock
\urldef\tempurl%
\url{https://doi.org/10.18653/v1/P17-2067}
\showDOI{\tempurl}


\bibitem[Wankhede et~al\mbox{.}(2018)]%
        {wankhede-2018}
\bibfield{author}{\bibinfo{person}{Shreyas Wankhede}, \bibinfo{person}{Ranjit
  Patil}, \bibinfo{person}{Sagar Sonawane}, {and}
  \bibinfo{person}{Prof.~Ashwini Save}.} \bibinfo{year}{2018}\natexlab{}.
\newblock \showarticletitle{Data Preprocessing for Efficient Sentimental
  Analysis}. In \bibinfo{booktitle}{\emph{2018 Second International Conference
  on Inventive Communication and Computational Technologies (ICICCT)}}.
  \bibinfo{pages}{723--726}.
\newblock
\urldef\tempurl%
\url{https://doi.org/10.1109/ICICCT.2018.8473277}
\showDOI{\tempurl}


\bibitem[Zannettou et~al\mbox{.}(2020)]%
        {zannettouMeasuringCharacterizingHate2020}
\bibfield{author}{\bibinfo{person}{Savvas Zannettou}, \bibinfo{person}{Mai
  Elsherief}, \bibinfo{person}{Elizabeth Belding}, \bibinfo{person}{Shirin
  Nilizadeh}, {and} \bibinfo{person}{Gianluca Stringhini}.}
  \bibinfo{year}{2020}\natexlab{}.
\newblock \showarticletitle{Measuring and {Characterizing} {Hate} {Speech} on
  {News} {Websites}}. In \bibinfo{booktitle}{\emph{12th {ACM} {Conference} on
  {Web} {Science}}} \emph{(\bibinfo{series}{{WebSci} '20})}.
  \bibinfo{publisher}{Association for Computing Machinery},
  \bibinfo{address}{New York, NY, USA}, \bibinfo{pages}{125--134}.
\newblock
\showISBNx{978-1-4503-7989-2}
\urldef\tempurl%
\url{https://doi.org/10.1145/3394231.3397902}
\showDOI{\tempurl}


\bibitem[Zubiaga et~al\mbox{.}(2017)]%
        {zubiagaExploitingContextRumour2017}
\bibfield{author}{\bibinfo{person}{Arkaitz Zubiaga}, \bibinfo{person}{Maria
  Liakata}, {and} \bibinfo{person}{Rob Procter}.}
  \bibinfo{year}{2017}\natexlab{}.
\newblock \showarticletitle{Exploiting {Context} for {Rumour} {Detection} in
  {Social} {Media}}.
\newblock In \bibinfo{booktitle}{\emph{Social {Informatics}}},
  \bibfield{editor}{\bibinfo{person}{Giovanni~Luca Ciampaglia},
  \bibinfo{person}{Afra Mashhadi}, {and} \bibinfo{person}{Taha Yasseri}}
  (Eds.). Vol.~\bibinfo{volume}{10539}. \bibinfo{publisher}{Springer
  International Publishing}, \bibinfo{address}{Cham},
  \bibinfo{pages}{109--123}.
\newblock
\showISBNx{978-3-319-67216-8 978-3-319-67217-5}
\urldef\tempurl%
\url{https://doi.org/10.1007/978-3-319-67217-5_8}
\showDOI{\tempurl}
\newblock
\shownote{Series Title: Lecture Notes in Computer Science}.


\end{thebibliography}






\end{document}